\documentclass{article}
\usepackage[hidelinks]{hyperref}
\usepackage{amsthm,amsmath,amssymb}
\usepackage{rotating}
\usepackage{titlesec}
\usepackage{mathrsfs}
\usepackage{multirow}
\RequirePackage{enumitem}
\RequirePackage{mathtools}
\usepackage[onehalfspacing]{setspace}
\usepackage{geometry}
\usepackage[super,comma]{natbib}
\usepackage{color, colortbl}
\definecolor{Gray}{gray}{0.9}

\numberwithin{equation}{section}
\theoremstyle{plain}

\titlelabel{\thetitle.\quad}

\newcommand{\beginsupplement}{%
    \setcounter{table}{0}
    \renewcommand{\thetable}{\arabic{table}}%
    \setcounter{figure}{0}
    \renewcommand{\thefigure}{\arabic{figure}}%
}

\begin{document}



\title{\textbf{Two-stage single-arm trials are rarely reported adequately}}
\author{\textbf{Michael J. Grayling\textsuperscript{1}, Adrian P. Mander\textsuperscript{2}}\\
\small 1. Population Health Sciences Institute, Newcastle University, Newcastle upon Tyne, UK\\
\small 2. Centre for Trials Research, Cardiff University, Cardiff, UK}
\date{}
\maketitle

\section*{Abstract}
\textbf{Purpose:} Two-stage single-arm trial designs are commonly used in phase II oncology to infer treatment effects for a binary primary outcome (e.g., tumour response). It is imperative that such studies be designed, analysed, and reported effectively. However, there is little available evidence on whether this is the case, particularly for key statistical considerations. We therefore comprehensively review such trials, examining in particular quality of reporting.\\
\textbf{Methods:} Published oncology trials that utilised ``Simon's two-stage design" over a 5 year period were identified and reviewed. Articles were evaluated on whether they reported sufficient design details, such as the required sample size, and analysis details, such as a confidence interval (CI). The articles that did not adjust their inference for the incorporation of an interim analysis were re-analysed to evaluate the impact on their reported point estimate and CI.\\
\textbf{Results:} Four hundred and twenty five articles that reported the results of a single treatment arm were included. Of these, 47.5\% provided the five components that ensure design reproducibility. Only 1.2\% and 2.1\% reported an adjusted point estimate or CI, respectively. Just 55.3\% of trials provided the final stage rejection bound, indicating many trials did not test the hypothesis the design is constructed to assess. Re-analysis of the trials suggests that reported point estimates underestimated treatment effects and that reported CIs were too narrow.\\
\textbf{Conclusion:} Key design details of two-stage single-arm trials are often unreported. Whilst inference is regularly performed, it is rarely done so in a way that removes the bias introduced by the interim analysis. In order to maximise their value, future studies must improve the way that they are analysed and reported.\\
\textbf{Keywords:} Adaptive design; Cancer; Estimation; Oncology; Phase II; Simon.

\section*{Introduction} \label{sec:int}

Whilst randomised trial designs are becoming more commonplace in phase II oncology\cite{grayling2019}, recent analyses indicate that single-arm trial designs remain the most widely utilised in this setting\cite{langrand2017}. Furthermore, the primary outcome variable is often dichotomous in phase II\cite{langrand2017}; typically chosen as objective response\cite{grayling2019} via RECIST\cite{eisenhauer2009}. Within the available class of single-arm trial designs for a binary primary outcome, what is commonly referred to as ``Simon's two-stage design"\cite{simon1989} is generally preferred\cite{ivanova2016}. The reasons for this preference are many, but principally Simon's proposal has been seen to provide a constructive means of formally testing for an efficacy signal via a simple design that requires only small sample size\cite{grayling2019}.

The habitual use of Simon's two-stage design has seen much research be conducted in to its effective utilisation. Recent work in this area includes methodology to account for deviation from the planned design\cite{belin2015,englert2015,zhao2015}, explorations of new criteria for simultaneously optimising the design and analysis\cite{bowden2012}, and evaluations of the value of such trials within wider phase II development plans\cite{grayling2016}. Indeed, an extensive list of publications have now addressed how to handle numerous issues that can arise in trials that use this design. Nonetheless, it is not known to what extent the advice provided in these publications has permeated through to practice.

Several authors have evaluated the reporting of phase II oncology trials without differentiating by design. For example, Grellety \textit{et al.}\cite{grellety2014} reviewed 156 phase II oncology trials published in 2011, assessing quality of reporting using two proposed scores. One of these, the Key Methodological Score (KMS), consisted of 3 items: provision of (i) a clear definition of a criterion of principal judgement, (ii) a clear justification of the number of patients included, and (iii) a clear definition of the population on which the principal/secondary judgement criteria were evaluated. They found that the median KMS was 2/3, whilst only 16.1\% of the analysed studies had a KMS of 3/3. Furthermore, Langrand-Escure \textit{et al.}\cite{langrand2017} reviewed 557 phase II and phase II/III oncology trials published in 2010-15 in three high-impact oncology journals, also appraising quality of reporting using the KMS. They concluded just 26.2\% of articles had a KMS of 3/3. They additionally found that a sample size calculation was missing in 66\% of the analysed articles.

These findings are concerning, but it is possible that they only scratch the surface of the issues in the use of two-stage single-arm designs in practice. In particular, to date, no paper has sought to ascertain the degree to which precise specific components of the design of such trials (e.g., the sample size required in each stage) is included in published reports. Moreover, no research has evaluated the frequency with which trialists have heeded the recommendations of the many articles that argue for the need for inference to be adjusted to account for the interim analysis. Finally, the extent to which deviation from the planned design occurs in practice, or the impact of this on type-I and type-II error-rates, is unknown. Given the extent of the phase II evidence base that comes from trials utilising two-stage single-arm designs, it is paramount that such studies be designed, analysed, and reported effectively. Therefore, with so little known about whether this is the case, we sought to systematically review a large number of trials that utilised Simon's two-stage design to ascertain issues in design, analysis, and reporting.

\section*{Methods} \label{sec:meth}

\subsection*{Two-stage group-sequential single-arm trial designs for a binary primary outcome}

We review phase II trials that used Simon's two-stage design. We therefore briefly summarise the statistical aspects of such trials.

The design evaluates a binary primary outcome, $x_i\in\{0,1\}$ from patient $i$, assumed to be distributed as $X_i \sim Bern(p)$. Thus, $p$ is the probability of success for the primary outcome. The following hypothesis is tested: $H_0 : p \le p_0$, with the type-I error-rate controlled to $\alpha\in(0,1)$ when $p=p_0$. The trial is powered to level $1-\beta\in(0,1)$ when $p=p_1>p_0$. Here, $p_0$ and $p_1$ are respectively commonly referred to as the maximal success probability that does not warrant further investigation and the minimal success probability that allows further investigation of the treatment. Often, $p_0$ is based on the historical success probability for the current standard-of-care.

The design includes a single interim analysis for futility (or a no-go decision) and is indexed by four values: $a_1, a, n_1$ and $n$. In stage 1, outcomes for $n_1$ patients are accumulated. If $s_{n_1} =\sum_{i=1}^{n_1}x_i \le a_1$ the trial terminates for futility with $H_0$ not rejected. Otherwise, outcomes for a further $n_2=n-n_1$ patients are gathered. Then, $H_0$ is rejected if $s_{n}=\sum_{i=1}^{n}x_i > a$ and not rejected otherwise. The design parameters, $a_1,a,n_1$, and $n$, are typically chosen as those that minimise some optimality criteria, amongst the combinations that meet the type-I error and power requirements. Simon\cite{simon1989} suggested two optimality criteria: (i) null-optimal, to minimise the expected sample size when $p=p_0$ and (ii) minimax, to minimise the maximal sample size $n$. Other optimality criteria have since been proposed\cite{hanfelt1999,jung2004a,mander2010}.

Post-trial inference could be performed using methods developed for one-sample proportions. For example, a confidence interval (CI) could be computed using the Clopper-Pearson\cite{clopper1934} approach. Depending on the stage of termination, a point estimate for $p$ could be given as $\hat{p}=s_{n_1}/n_1$ or $s_{n}/n$. However, it is well known that the inclusion of an interim analysis necessitates that adjusted inference be performed in order to compute $p$-values that are consistent with the decision on $H_0$, to acquire CIs with the desired coverage, and to reduce the bias in the point estimate\cite{porcher2012}. Many such adjusted methods have been proposed, including that of Jung \textit{et al.}\cite{jung2006} for $p$-values, Jennison and Turnbull\cite{jennison1983} for CIs, and Jung and Kim\cite{jung2004b} for point estimates. A selection of methods for handling deviation from the planned design (i.e., scenarios in which the interim or final analysis is conducted with a sample size different from $n_1$ or $n$ respectively) have also been developed\cite{belin2015,englert2015,zhao2015,wu2008}. We provide details on all methods used later in the Supplementary Materials.

\subsection*{Literature review}

Further details on the literature review are included in the Supplementary Materials. Key points are given here.

\subsubsection*{Inclusion criteria}

To identify articles for potential inclusion, PubMed was searched on February 21 2018 using the following term: (``2013/01/01"[Date - Publication] : ``2017/12/31"[Date - Publication]) AND Clinical Trial[Publication Type] AND (phase II[Title/Abstract] OR phase 2[Title/Abstract]) AND (cancer[All Fields] OR oncology[All Fields]). This returned 5344 articles for review.

The key inclusion criteria were: (i) full-length articles (i.e., no short communications), (ii) primary publications on a trial's complete results (i.e., no secondary or preliminary analyses), and (iii) reports results for at least one treatment arm that was designed and analysed using ``Simon's two-stage design".

Five hundred and thirty-four of the 5344 returned articles (10.0\%) were randomly selected for evaluation for inclusion by MJG and APM. The authors agreed on inclusion for 520 articles (97.3\%). Given the high-level of agreement and the low number of reasons for disagreement, the remaining articles were assessed for inclusion by MJG only, with discussion with APM where required.

\subsubsection*{Data extraction}

Data on each of the questions listed in Table~\ref{tab:qu} was extracted by MJG for each arm, in each article, deemed eligible for inclusion. To establish the reliability of this extraction, data extracted by MJG was compared with that independently extracted on 58 arms by APM. Across 14 questions requiring non-binary value extraction (e.g., `Q5. What was the value of $p_0$?') the duplicate extractions agreed 96.2\% of the time. Across a wider set of 26 questions, including those requiring only binary value extraction, the duplicate extractions agreed 94.3\% of the time. The final extracted data is reported as percentages or through figures as appropriate.

\begin{table}[htbp]
	\begin{center}
		\caption{The questions for which data was extracted for all included treatment arms.\label{tab:qu}}
		\begin{tabular}{ll}
			\hline
			Question & Statement \\
			\hline
	\rowcolor{Gray}
			Q1 & How many included treatment arms are reported in this article?\\
			Q2 & What type of cancer was the article focused on?\\
				\rowcolor{Gray}
			Q3 & Does the article use the phase ``Simon two-stage", or similar, or cite Simon (1989)\cite{simon1989}?\\
			Q4 & What was the chosen design optimality criteria?\\
				\rowcolor{Gray}
			Q5 & What was the value of $p_0$?\\
			Q6 & Was a justification given for the value of $p_0$?\\
				\rowcolor{Gray}
			Q7 & What was the value of $p_1$?\\
			Q8 & What was the value of $\alpha$?\\
				\rowcolor{Gray}
			Q9 & What was the value of $\beta$?\\
			Q10 & What was the value of $a_1$?\\
				\rowcolor{Gray}
			Q11 & What was the value of $a$?\\
			Q12 & What was the value of $n_1$?\\
				\rowcolor{Gray}
			Q13 & What was the value of $n$?\\
			Q14 & Was a target recruitment goal larger than $n$ stated?\\
				\rowcolor{Gray}
			Q15 & What stage did the trial terminate?\\
			Q16 & If `2' to Q15, did they explicitly state the criteria for progression to stage 2 was met?\\
				\rowcolor{Gray}
			Q17 & If `2' to Q15, what was the realised sample size in stage 1?\\
			Q18 & If `2' to Q15, what was the number of successes in stage 1?\\
				\rowcolor{Gray}
			Q19 & Did the stage of termination have to be inferred from the enrolled/analysed sample size?\\
			Q20 & What was the total enrolled sample size?\\
				\rowcolor{Gray}
			Q21 & Did they report a point estimate, $p$-value, or confidence interval?\\
			Q22 & If `Yes' to Q21, what was the sample size assumed in the analysis?\\
				\rowcolor{Gray}
			Q23 & If `Yes' to Q21, what was the total number of successes assumed in the analysis?\\
			Q24 & If `Yes' to Q21, did they report a point estimate?\\
				\rowcolor{Gray}
			Q25 & If `Yes' to Q24, did they state they used an adjusted point estimate?\\
			Q26 & If `Yes' to Q24, what was the reported point estimate?\\
				\rowcolor{Gray}
			Q27 & If `Yes' to Q21, did they report a $p$-value?\\
			Q28 & If `Yes' to Q27, did they state they used an adjusted $p$-value?\\
				\rowcolor{Gray}
			Q29 & If `Yes' to Q27, what was the reported $p$-value?\\
			Q30 & If `Yes' to Q21, did they report a confidence interval?\\
				\rowcolor{Gray}
			Q31 & If `Yes' to Q30, did they state they used an adjusted confidence interval?\\
			Q32 & If `Yes' to Q30, what confidence interval level was used?\\
				\rowcolor{Gray}
			Q33 & If `Yes' to Q30, what was the reported confidence interval's lower limit?\\
			Q34 & If `Yes' to Q30, what was the reported confidence interval's upper limit?\\
			\hline
		\end{tabular}
	\end{center}
\end{table}

\subsubsection*{Trial re-analyses}

Given absence of evidence is not evidence of absence, included articles that did not state they reported an adjusted point estimate (Q25) were re-analysed where possible (i.e., subject to reporting the required design components). This enabled evaluation of which of seven possible point estimates (the unadjusted estimate and six adjusted estimates) the reported point estimate (Q26) was consistent with, to the reported number of decimal places.

Equivalent computations were conducted for those articles that did not state they reported an adjusted CI (Q30); the reported CI (Q32-34) was compared for its consistency with four unadjusted and two adjusted CIs.

The re-analyses were limited to those trials (i) adjudged to have terminated in stage 2, as point estimate and CI procedures do not in general adjust when a trial terminates in stage 1 and (ii) that reported the number of successes and sample size assumed in the analysis, as these are required to calculate unadjusted point estimates and CIs. To re-analyse using the adjusted inferential procedures, $a_1$ and $n_1$ were additionally required to have been reported clearly, as these are requisite to adjusted inference. 

\section*{Results}\label{sec:res}

\subsection*{Included articles}

Five hundred articles were deemed eligible for inclusion. Four hundred and twenty-five of these reported the results of a single eligible treatment arm. The remaining 75 articles reported results for an additional 204 eligible treatment arms (arms per article: median 2, range [2,15]).

To remove the need to account for skew caused by the quality of the articles reporting multiple included treatment arms, we discuss here the findings for only the 425 included articles that reported the results of a single eligible treatment arm. Findings for the remaining 75 articles are given in the Supplementary Materials.

Table~\ref{tab:desc} provides descriptors on the 425 articles. At least 15.8\% of the articles came from each allowed publication year, with a decline in inclusion by year potentially indicating decreasing use of the considered design type. The included articles were published in 100 distinct journals, with more than ten articles included for nine journals. The type of cancer under consideration varied widely, though lung and lymph cancers together accounted for 26.4\% of the articles.

One hundred and ten trials (25.9\%) were judged to have terminated in stage 1, in contrast to 298 in stage 2 (70.1\%). Amongst the 298 judged to have terminated in stage 2, only 80 (26.4\%) stated the criteria had been met for progression to stage 2; indicating that this judgement often had to be based on the enrolled and analysed sample sizes. For 17 articles (4.0\%) it was not possible to ascertain when the trial terminated; this was caused by the final sample size being between $n_1$ and $n$, or because neither of the planned stage-wise sample sizes were reported.

\begin{table}[htbp]
	\begin{center}
		\caption{Descriptors on the 425 included articles that reported the results of a single eligible treatment arm. The denominator for computing percentages (given to 1 decimal place) is 425 in all instances.\label{tab:desc}}
		\begin{tabular}{llr}
			\hline
			Descriptor & Value & Number (\%) \\
			\hline
				\rowcolor{Gray}
			Publication year & 2013 & 102 (24.0) \\
			\cellcolor{Gray}                 & 2014 & 101 (23.8) \\
			                 	\rowcolor{Gray}
			                 & 2015 & 79 (18.6) \\
			\cellcolor{Gray}                 & 2016 & 76 (17.9) \\
			                 	\rowcolor{Gray}
			                 & 2017 & 67 (15.8) \\
			Journal & \textit{Cancer Chemother Pharmacol} & 44 (10.4) \\
			        & \cellcolor{Gray}\textit{Ann Oncol}                  & \cellcolor{Gray} 30 (7.1) \\
			        & \textit{Invest New Drugs}           & 23 (5.4) \\
			        & \cellcolor{Gray}\textit{Lung Cancer}                & \cellcolor{Gray}17 (4.0) \\
			        & \textit{Cancer}                     & 15 (3.5) \\
			        & \cellcolor{Gray}\textit{BMC Cancer}                 & \cellcolor{Gray}13 (3.1) \\
			        & \textit{J Clin Oncol}               & 13 (3.1) \\
			        & \cellcolor{Gray}\textit{Br J Haematol}              & \cellcolor{Gray}11 (2.6) \\
			        & \textit{Lancet Oncol}               & 11 (2.6) \\
			        & \cellcolor{Gray}Other (91 journals)                      & \cellcolor{Gray}248 (58.4) \\
			\cellcolor{Gray}Cancer & Lung          & 59 (13.9) \\
			\rowcolor{Gray}
			       & Lymph         & 53 (12.5) \\
			      \cellcolor{Gray} & Colon/rectum  & 32 (7.5) \\
			       \rowcolor{Gray}
			       & Breast        & 28 (6.6) \\
			      \cellcolor{Gray} & Stomach       & 25 (5.9) \\
			       \rowcolor{Gray}
			       & Head and neck & 23 (5.4) \\
			      \cellcolor{Gray} & Blood         & 22 (5.2) \\
			       \rowcolor{Gray}
			       & Kidney        & 21 (4.9) \\
			      \cellcolor{Gray} & Other         & 162 (38.1) \\
			Stage of termination & \cellcolor{Gray}1 & \cellcolor{Gray}110 (25.9)\\
			& 2: Stated the criteria had been met for progression & 80 (18.8)\\
			& \cellcolor{Gray}2: Did not state the criteria had been met for progression & \cellcolor{Gray}218 (51.3)\\
			& Unclear & 17 (4.0)\\
			\hline
		\end{tabular}
	\end{center}
	\label{tab:1}
\end{table}

\subsection*{Reporting of design characteristics}

Extracted data on the reporting of design characteristics is summarised in Table~\ref{tab:des}. Whilst 380 articles (89.4\%) clearly stated $p_0$, only 78 (18.4\%) provided a justification for this value. The success probability $p_1$ was often reported (391 articles; 92.0\%), as were the desired type-I (372 articles; 87.5\%) and type-II error-rates (382 articles; 89.9\%). The chosen optimality criteria was stated in only 240 articles (56.5\%). This is the principal driver of the fact that only 202 articles (47.5\%) reported $p_0$, $p_1$, $\alpha$, $\beta$, and the optimality criteria; the five components that ensure a design can be easily reproduced.

Whilst $a_1$ (349 articles; 82.1\%), $n_1$ (371 articles; 87.3\%), and $n$ (394 articles; 92.7\%) were all regularly reported, $a$ was given in only 235 articles (55.3\%). This drives the result that just 221 articles (52.0\%) reported the four components $a_1$, $a$, $n_1$, and $n$; those that enable a design's operating characteristics to be computed for any $p$.

\begin{figure}
	\centering
	\includegraphics[width=0.75\textwidth]{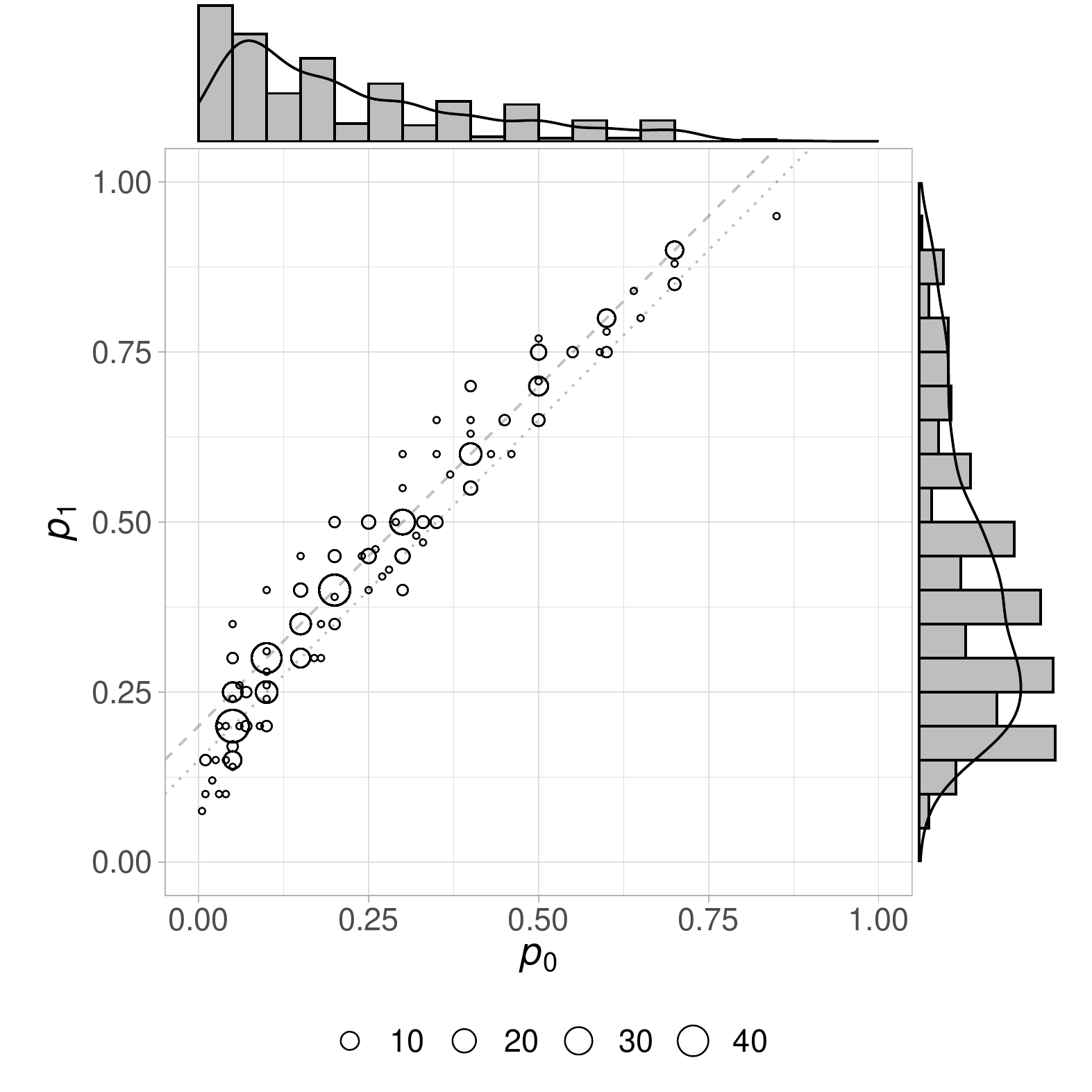}
	\caption{The combinations of $p_0$ and $p_1$ given in the 373/425 articles (87.8\%) that reported both of these components is depicted. The size of the circles indicates the number of articles that had a particular combination. Densigrams are given on the axes to display the marginal distributions of $p_0$ and $p_1$. The dotted line indicates the lower quartile (0.15) of the distribution of $p_1-p_0$; the dashed line indicates the median and upper quartile (0.2) of this distribution.\label{fig1}}
\end{figure}

Figure~\ref{fig1} depicts the values for $p_0$ and $p_1$ given in the 373 articles (87.8\%) that reported both of these quantities. The median value of $p_1-p_0$ was 0.2.

\begin{table}[htbp]
	\begin{center}
		\caption{Reporting of the design of the 425 included articles that reported the results of a single eligible treatment arm. The denominator for computing percentages (given to 1 decimal place) is 425 in all instances.\label{tab:des}}
		\begin{tabular}{lr}
			\hline
			Criteria & Number (\%) \\
			\hline
			\rowcolor{Gray}
			Used the phrase ``Simon two-stage" (or similar) or cited Simon (1989)\cite{simon1989}                & 357 (84.0) \\
			Clearly stated $p_0$                                                                                & 380 (89.4) \\
			\rowcolor{Gray}
			Gave a justification for $p_0$                                                                      & 78 (18.4) \\
			\hspace{25pt}Citation given                                                                         & 40 (9.4) \\
			\rowcolor{Gray}
			\hspace{25pt}Justification given but no citation                                                    & 38 (8.9) \\
			Clearly stated $p_1$                                                                                & 391 (92.0) \\
			\rowcolor{Gray}
			Clearly stated $\alpha$                                                                             & 372 (87.5) \\
			\hspace{25pt}$\alpha=0.05$                                                                          & 231 (54.4) \\
			\rowcolor{Gray}
			\hspace{25pt}$\alpha=0.1$                                                                           & 103 (24.2) \\
			Clearly stated $\beta$                                                                              & 382 (89.9) \\
			\rowcolor{Gray}
			\hspace{25pt}$\beta=0.1$                                                                            & 165 (38.8) \\
			\hspace{25pt}$\beta=0.2$                                                                            & 173 (40.7) \\
			\rowcolor{Gray}
			Clearly stated the optimality criteria                                                              & 240 (56.5) \\
			\hspace{25pt}Null-optimal                                                                           & 142 (33.4) \\
			\rowcolor{Gray}
			\hspace{25pt}Minimax                                                                                & 93 (21.9) \\
			\hspace{25pt}Admissable                                                                             & 4 (0.9) \\
			\rowcolor{Gray}
			\hspace{25pt}Other                                                                                  & 1 (0.2) \\
			Clearly stated $a_1$                                                                                & 349 (82.1) \\
			\rowcolor{Gray}
			Clearly stated $a$                                                                                  & 235 (55.3) \\
			Clearly stated $n_1$                                                                                & 371 (87.3) \\
			\rowcolor{Gray}
			Clearly stated $n$                                                                                  & 394 (92.7) \\
			Indicated the recruitment target was greater than $n$                                               & 117 (27.5) \\
			\rowcolor{Gray}
			Clearly stated $p_0$ and $p_1$                                                                      & 373 (87.8) \\
			Clearly stated $p_0$, $p_1$, $\alpha$, and $\beta$                                                  & 340 (80.0) \\
			\rowcolor{Gray}
			Clearly stated $a_1$, $a$, $n_1$, and $n$                                                           & 221 (52.0) \\
			Clearly stated $p_0$, $p_1$, $\alpha$, $\beta$, and the optimality criteria                         & 202 (47.5) \\
			\rowcolor{Gray}
			Clearly stated $p_0$, $p_1$, $\alpha$, $\beta$, the optimality criteria, $a_1$, $a$, $n_1$, and $n$ & 109 (25.6) \\
			\hline
		\end{tabular}
	\end{center}
\end{table}

\subsection*{Reporting of inferential procedures}

Extracted data on the reporting of the inferential procedures performed in the 425 included articles is summarised in Table~\ref{tab:inf}, with additional stratification by stage of termination.

In total, 375 articles (88.2\%) reported either a point estimate, $p$-value, or CI for their primary outcome. This figure was larger for the trial's adjudged to have terminated in stage 2 (287/298 articles; 96.3\%).

Whilst point estimates were often reported (372 articles; 87.8\%), only 5 articles (1.2\%) stated that they had reported an adjusted point estimate. In contrast, $p$-values were rarely reported (4 articles; 1.3\%). For CIs, just 233 articles (54.8\%) reported any type of CI, with only 9 (2.1\%) indicating that they reported an adjusted CI. 

To evaluate whether the articles that reported a point estimate or CI but did not indicate it was adjusted were consistent (to their reported number of decimal places) with unadjusted or adjusted analyses, the trials were re-analysed (Table~\ref{tab:rean}). Two hundred and seventy (96.1\%) of the re-analysed articles reported point estimates consistent with an unadjusted estimate. However, 133/228 articles (58.3\%) for which adjusted point estimates could be calculated were also consistent with at least one adjusted estimate. For the CIs, 116/178 articles (65.2\%) that were re-analysed were consistent with at least one unadjusted interval. Far fewer articles (3/140; 2.1\%) for which adjusted intervals could be computed were consistent with at least one adjusted interval.

\begin{figure}
	\centering
	\includegraphics[width=0.75\textwidth]{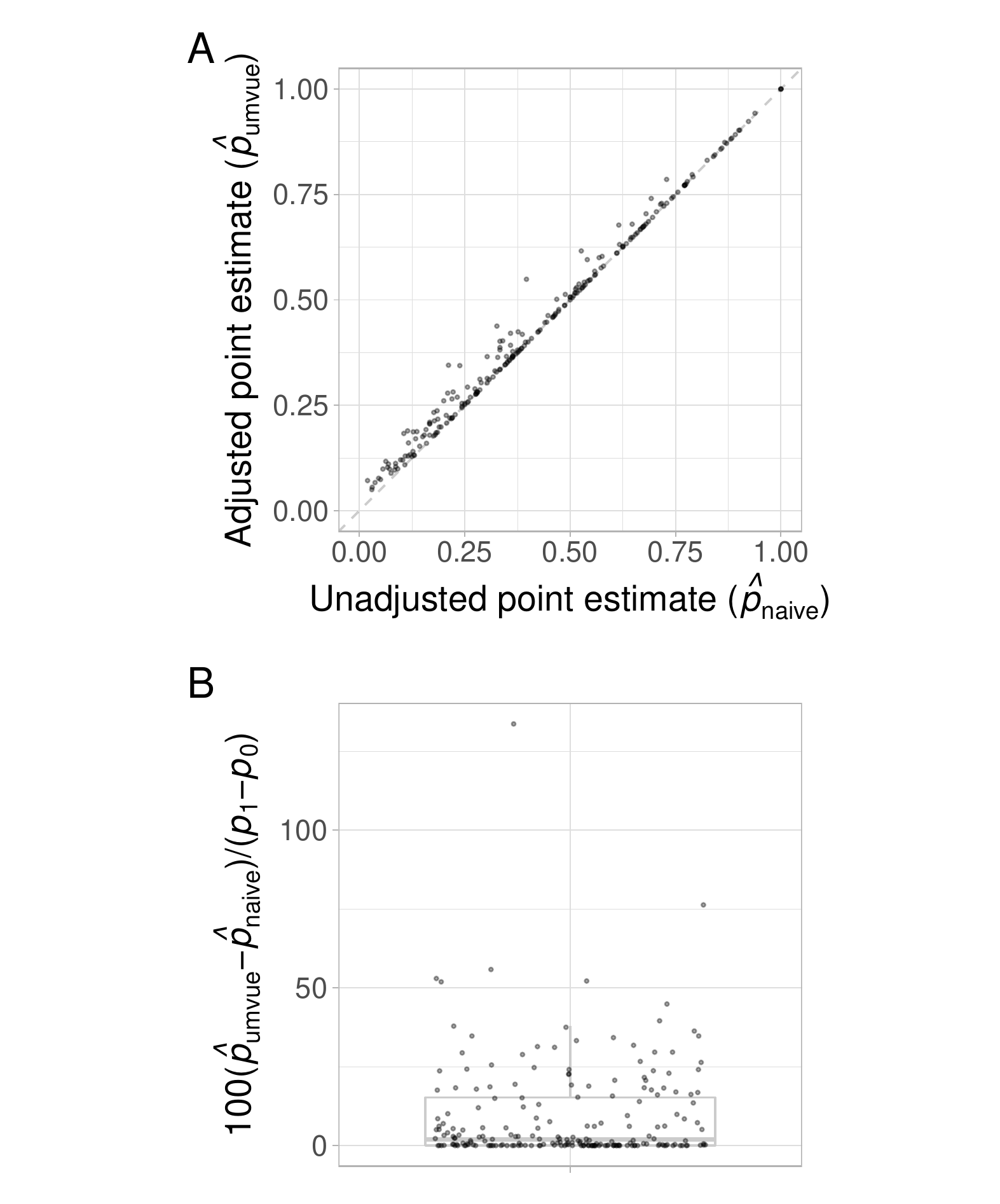}
	\caption{In A, a comparison of the naive unadjusted point estimate ($\hat{p}_\text{naive}$) and the uniform minimum variance unbiased estimate (UMVUE, $\hat{p}_\text{umvue}$) is given for the 233 trials that terminated in stage 2 where the UMVUE could be computed. In B, the difference between $\hat{p}_\text{naive}$ and $\hat{p}_\text{umvue}$ is presented as a percentage of $p_1-p_0$, along with a boxplot to indicate the distribution of this data.\label{fig2}}
\end{figure}

To visualise the impact of not utilising adjusted inferential procedures, Figure~\ref{fig2}A displays the unadjusted estimate ($\hat{p}_\text{naive}$) against the uniform minimum variance unbiased estimate\cite{jung2004b} (UMVUE, $\hat{p}_\text{umvue}$) for the 233 trials that terminated in stage 2 where the UMVUE could be computed. The difference between the unadjusted estimate and the UMVUE is presented as a percentage of $p_1-p_0$ in Figure~\ref{fig2}B. Together, these plots indicate that whilst the difference between the unadjusted and adjusted estimates may often be small, there are instances in which it is large; in 25 cases more than 25\% of the difference $p_1-p_0$.

\begin{figure}
	\centering
	\includegraphics[width=0.75\textwidth]{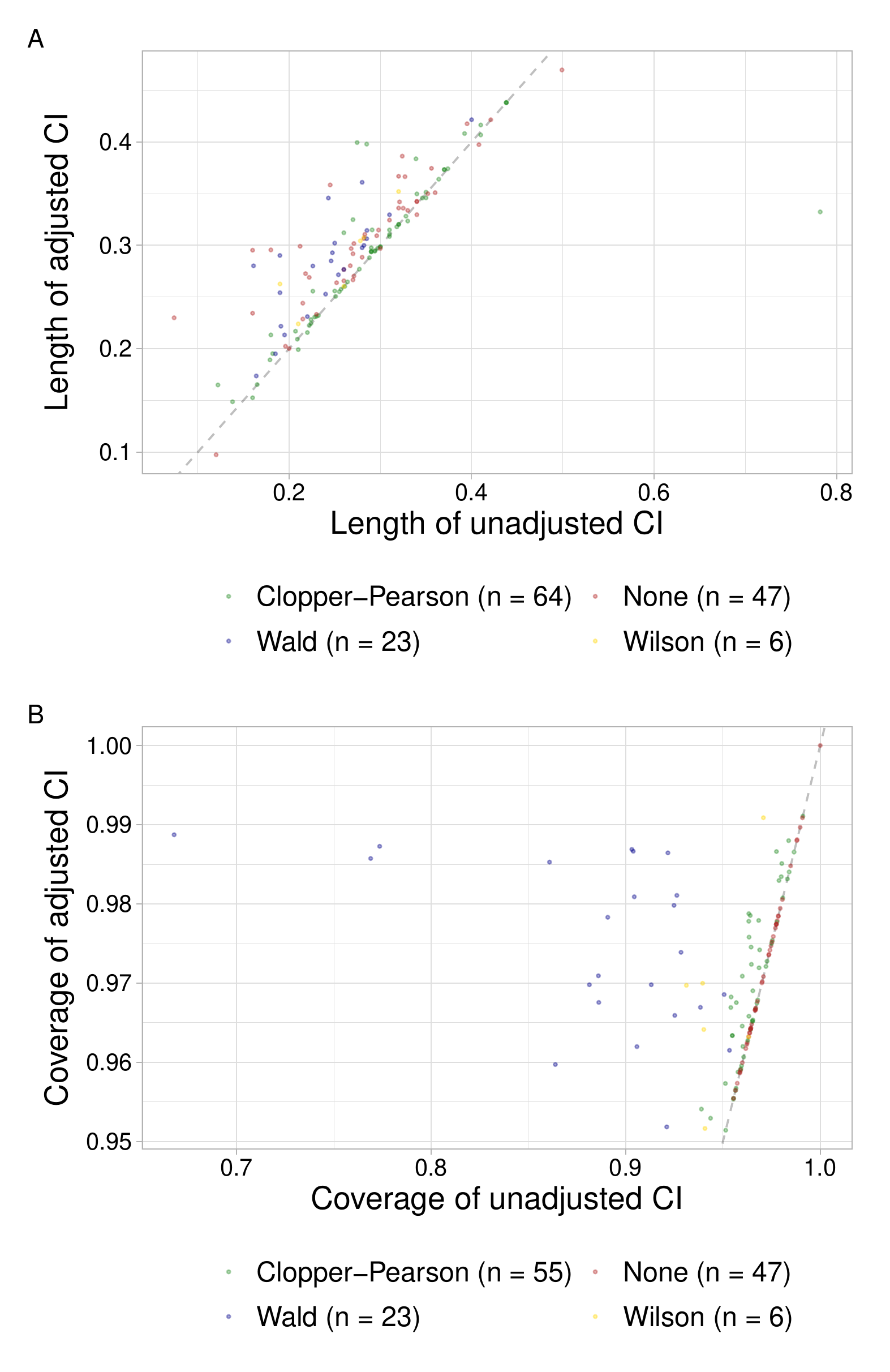}
	\caption{In A, the length of the reported unadjusted CI is compared to the length of the corresponding adjusted CI proposed by Jennison and Turnbull\cite{jennison1983} for the 140 articles for which this adjusted CI could be computed. In B, the respective coverage of these unadjusted and adjusted CIs when $p=\hat{p}_\text{umvue}$ is given for the 131 of these articles in which the target coverage was 0.95. In both cases, points are coloured by the unadjusted CI that the re-analysis indicated the reported CI matched with. For those CIs that matched none of the unadjusted CIs, Clopper-Pearson\cite{clopper1934} was used to compute the coverage.\label{fig3}}
\end{figure}

Similar visualisations are provided in Figure~\ref{fig3}. Figure~\ref{fig3}A displays the length of the reported unadjusted CI against the length of the corresponding adjusted CI proposed by Jennison and Turnbull\cite{jennison1983} for the 140 trials for which this adjusted CI could be computed. Figure~\ref{fig3}B compares the respective coverage of these unadjusted and adjusted CIs when $p=\hat{p}_\text{umvue}$ for the 131 trials in which the target coverage was 0.95. In general, the length of the unadjusted CI is shorter than the corresponding adjusted CI, which is reflected in the coverage being below the desired level for the unadjusted procedure in several instances.

\begin{table}[htbp]
	\begin{center}
		\caption{Reporting of inferential procedures performed in the 425 included articles that reported the results of a single eligible treatment arm, with additional stratification by stage of termination. The denominators for computing percentages (given to 1 decimal place) in the three columns are 110, 298, and 425 respectively unless stated otherwise.\label{tab:inf}}
		\begin{tabular}{lrrr}
			\hline
			Criteria & Stage 1 & Stage 2 & All\\
			& Number (\%) & Number (\%) & Number (\%)\\
			\hline
			\rowcolor{Gray}
			Reported a point estimate, $p$-value, or confidence & 72 (65.5)    & 287 (96.3)    & 375 (88.2) \\
			\rowcolor{Gray}
			interval for the primary outcome & & &\\
			Reported a point estimate & 70 (63.6)    & 287 (96.3)    & 372 (87.8) \\
			\rowcolor{Gray}
			Stated the point estimate had been adjusted for & 0 (0)        & 5 (1.7)       & 5 (1.2) \\
			\rowcolor{Gray}
			the two-stage design & & &\\
			Reported a $p$-value                                                                      & 0 (0)        & 4 (1.3)       & 4 (0.9) \\
			\rowcolor{Gray}
			Stated the $p$-value had been adjusted for the & 0 (0)        & 3 (1.0)       & 3 (0.7) \\
			\rowcolor{Gray}
			two-stage design & & &\\
			Reported a confidence interval                                                          & 40 (36.4)    & 187 (62.8)    & 233 (54.8) \\
			\rowcolor{Gray}
			Stated the confidence interval had been & 0 (0)        & 9 (3.0)       & 9 (2.1) \\
			\rowcolor{Gray}
			adjusted for the two-stage design & & &\\
			Analysis performed assuming a sample equal & 27/70 (38.6) & 72/278 (25.9) & 99/348 (28.4) \\
			to that given in the design & & &\\
			\hline
		\end{tabular}
	\end{center}
\end{table}

\begin{table}[htbp]
	\begin{center}
		\caption{Re-analysis of the subset of the 425 articles that reported a point estimate or confidence interval not stated to have been adjusted. Consistency is measured in all cases against the reported number of decimal places. The denominators for computing percentages (given to 1 decimal place) are given in each row. Note that the re-analysis is limited to those articles that were judged to have terminated in stage 2.\label{tab:rean}}
		\begin{tabular}{lrrr}
			\hline
			Criteria & Number (\%) \\
			\hline
			\rowcolor{Gray}
			Reported a point estimate not stated as adjusted and clearly reported the & 281/298 (94.2)\\
			\rowcolor{Gray}
			number of successes and sample size assumed in the analysis & \\
			Reported point estimate consistent with an unadjusted estimate & 270/281 (96.1)\\
			\rowcolor{Gray}
			Reported point estimate consistent with at least one adjusted estimate & 133/228 (58.3)\\
			Reported a confidence interval not stated as adjusted and clearly reported its &  178/298 (59.7)\\
			level, the number of successes, and sample size assumed in the analysis & \\
			\rowcolor{Gray}
			Reported confidence interval consistent with at least one unadjusted interval & 116/178 (65.2)\\
			Reported confidence interval consistent with at least one adjusted interval & 3/140 (2.1)\\
			\hline
		\end{tabular}
	\end{center}
\end{table}

\begin{figure}
	\centering
	\includegraphics[width=0.75\textwidth]{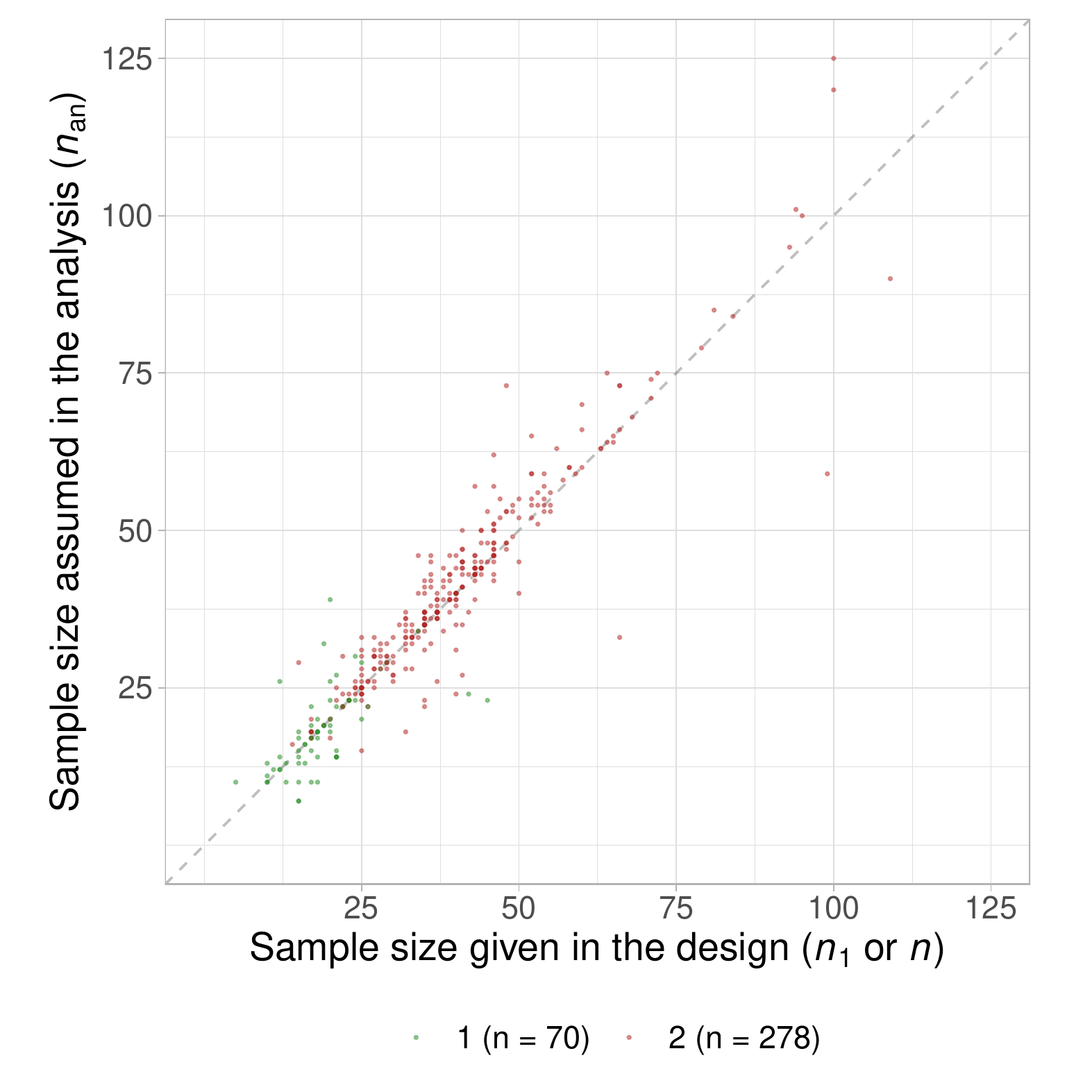}
	\caption{A comparison of the sample size required by the design and that assumed in the analysis for 348 articles. Colour indicates the judged stage of trial termination.\label{fig4}}
\end{figure}

Finally, note that 348 trials that were judged to have ended in stage 1 or stage 2 reported a point estimate, $p$-value, or CI, as well as the sample size required by their design. Amongst these, just 99 (28.4\%) performed their analysis using the planned sample size. Differences in the planned and analysed sample sizes are shown in Figure~\ref{fig4}.

\section*{Discussion} \label{sec:disc}

A large proportion of all phase II evidence comes from trials conducted using Simon's two-stage design, as exemplified by the number of articles found to be eligible in this review. This necessitates that such studies be designed, analysed, and reporting effectively. We evaluated the degree to which this is true through a comprehensive review of trials conducted over a 5 year period.

It is easy to argue that the reporting of design components was poor. In particular, the reproducibility of determined designs is limited by infrequent reporting of $p_0$, $p_1$, $\alpha$, and $\beta$ in unison. It is also alarming that only 18.4\% of trials provided a justification for $p_0$, considering the interpretation of study results is highly dependent on this value. Furthermore, it may be considered disappointing that so many trials chose standard target error-rates (e.g, $\alpha=0.05$), as it has been highlighted that small concessions in this regard can lead to notable efficiency gains\cite{khan2012}. Similar statements are true of the chosen optimality criteria, with previous work noting that the often-used minimax and null-optimal designs may routinely not be the most advisable choice\cite{jung2004a,mander2012}.

Few articles stated that they utilised adjusted inferential procedures. Given there is no additional cost to using these methods, this is a disappointing finding. Figures~\ref{fig3} and~\ref{fig4} together indicate that the result of this in practice may be that phase II trials utilising Simon's two-stage design are conservative in their reported point estimate, but anti-conservative in the width of their CI. It is also surprising that only 54.8\% of articles included a CI of any kind. The size of single-arm trials makes the uncertainty around a point estimate important to quantify; we encourage future studies to include such details.

Many final analyses were performed with a sample size different from that specified in the design (71.6\%); this is not surprising given many trials indicated they planned to over-recruit to allow for attrition (Table~\ref{tab:des}). This highlights the need for trialists to plan for such design deviation, and echoes previous findings from Koyama and Chen\cite{koyama2008}. We initially hoped to extract data on how trials handled design deviation when interpreting their results. This was ultimately judged to be too subjective an endeavour to complete, as many studies interpreted their findings through informal comparison of their point estimate or CI bounds to $p_0$ and/or $p_1$.

\begin{figure}
	\centering
	\includegraphics[width=0.75\textwidth]{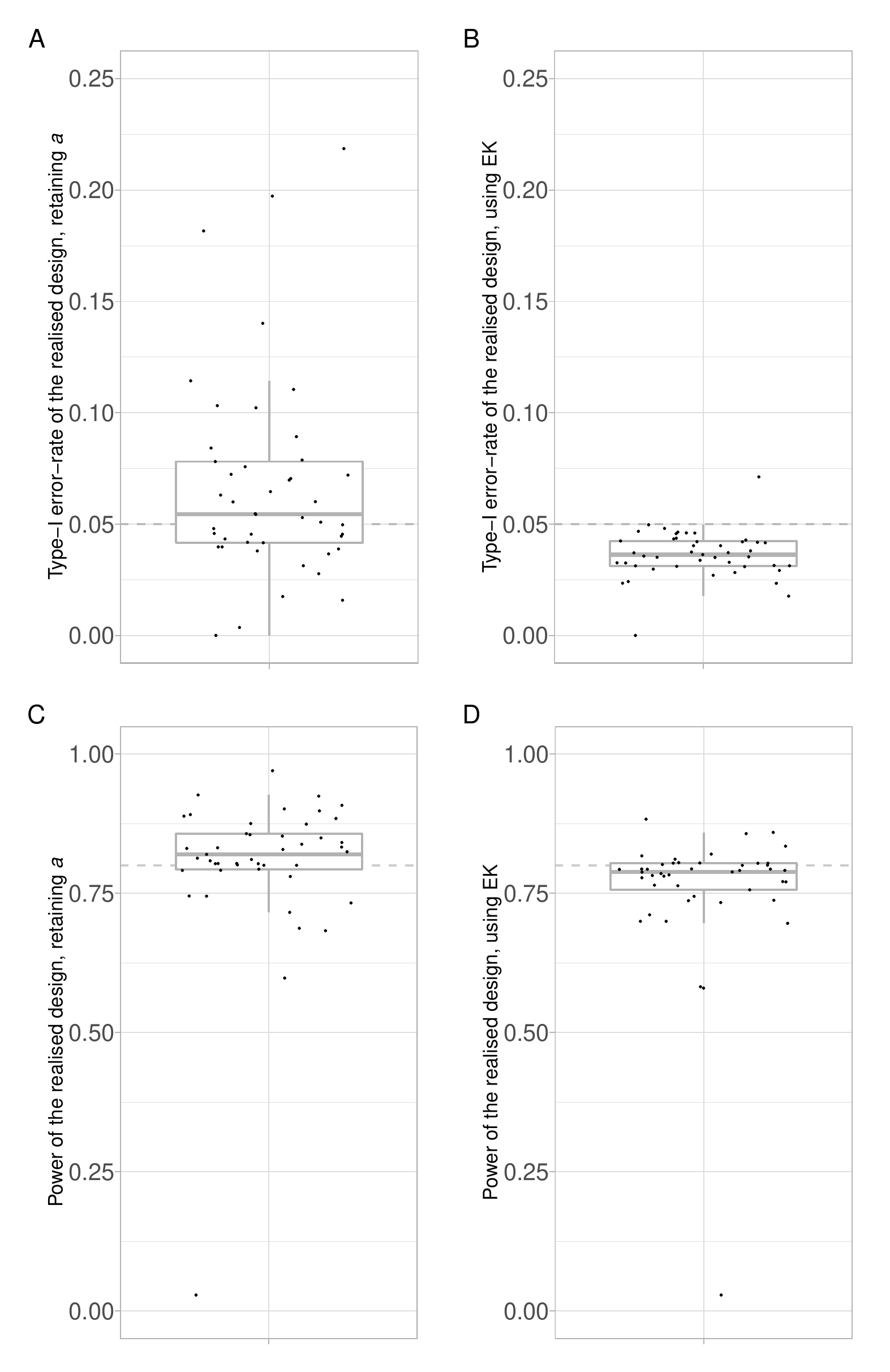}
	\caption{A comparison of the type-I error-rate and power of 45 trials that terminated in stage 2 and specified $\alpha=0.05$ and $1-\beta=0.8$, is given. In A and C, results are given for when the reported value of $a$ is retained with the sample size assumed in the analysis. In B and D, results are given when the approach proposed by Englert and Kieser\cite{englert2015} (EK) is implemented to account for deviation. Note that in B, one trial does not attain the target type-I error-rate; this is a consequence of the original reported design also not attaining the desired error-rate, and not an inadequacy of the methodology to account for deviation.\label{fig5}}
\end{figure}

The difficulties in practice of attaining the planned sample size may be reflected in the fact that only 55.3\% of trials reported $a$. This lack of reporting of $a$ also indicates that many trials that utilise Simon's two-stage design do not formally test the hypothesis they claim to. We note that methodology to comprehensively handle over- and under-running is available. Its use is depicted in Figure~\ref{fig5}, which provides the error-rates for 45 trials when the methodology of Englert and Kieser\cite{englert2015} is implemented. Using this methodology, trials are assured to conform to their desired type-I error-rate, and it appears sample sizes that enable power to reach close to the desired level may have been achieved in practice. As a contrast, the error-rates if $a$ was retained at the final analysis are also shown, while other possible methods of interpreting trial results are given in the Supplementary Materials. We note that without utilising established methodology to specifically account for design deviation, many trials may be interpreting their findings in a manner associated with a high probability of erroneous decision making.

We acknowledge a number of limitations to our work. Firstly, only a 10\% duplicate extraction was performed. Given the strength of our findings, though, we note that it is unlikely our conclusions would be altered by additional duplicate extractions. It is also impossible to be certain that those trials that did not state they utilised an adjusted inferential procedure had used an unadjusted method. However, our re-analyses (Table~\ref{tab:rean}) do provide evidence that this may be the case.

In all, in light of work that has assessed whether reported randomised trials conform to the recommendations of the CONSORT guidance\cite{turner2012}, our findings should perhaps not be surprising. Nonetheless, it may have been hoped that the simplicity of Simon's design would lead to effective reporting. Our results indicate that a CONSORT extension specific to the requirements of single-arm oncology trials may be warranted.

\bibliographystyle{vancouver}
\bibliography{grayling_bibliography}

\begin{thebibliography}{10}

\bibitem{grayling2019}
Grayling M, Dimairo M, Mander A, Jaki T.
\newblock A review of perspectives on the use of randomization in phase II
  oncology trials.
\newblock \emph{JNCI - J Natl Cancer Inst}  2019;111:1255--62.

\bibitem{langrand2017}
Langrand-Escure J, Rivoirard R, Oriol M, et~al.
\newblock Quality of reporting in oncology phase II trials: A 5-year assessment
  through systematic review.
\newblock \emph{PLoS One}  2017;12:e0185536.

\bibitem{eisenhauer2009}
Eisenhauer E, Therasse P, Bogaerts J, et~al.
\newblock New response evaluation criteria in solid tumours: Revised RECIST
  guideline (version 1.1).
\newblock \emph{Eur J Cancer}  2009;45:228--47.

\bibitem{simon1989}
Simon R.
\newblock Optimal two-stage designs for phase II clinical trials.
\newblock \emph{Control Clin Trials}  1989;10:1--10.

\bibitem{ivanova2016}
Ivanova A, Paul B, Marchenko O, Song G, Patel N, Moschos S.
\newblock Nine-year change in statistical design, profile, and success rates of
  phase II oncology trials.
\newblock \emph{J Biopharm Stat}  2016;26:141--9.

\bibitem{belin2015}
Belin L, Broet P, De~Rycke Y.
\newblock A rescue strategy for handling unevaluable patients in Simon's two
  stage design.
\newblock \emph{PLoS One}  2015;10:e0137586.

\bibitem{englert2015}
Englert S, Kieser M.
\newblock Methods for proper handling of overrunning and underrunning in phase
  II designs for oncology trials.
\newblock \emph{Stat Med}  2015;34:2128--37.

\bibitem{zhao2015}
Zhao J, Yu M, Feng X.
\newblock Statistical inference for extended or shortened phase II studies
  based on Simon's two-stage designs.
\newblock \emph{BMC Med Res Methodol}  2015;15:48.

\bibitem{bowden2012}
Bowden J, Wason J.
\newblock Identifying combined design and analysis procedures in two‐stage
  trials with a binary end point.
\newblock \emph{Stat Med}  2012;31:3874--84.

\bibitem{grayling2016}
Grayling M, Mander A.
\newblock Do single‐arm trials have a role in drug development plans
  incorporating randomised trials?
\newblock \emph{Pharm Stat}  2016;15:143--51.

\bibitem{grellety2014}
Grellety T, Petit-Moneger A, Diallo A, Mathoulin-Pelissier S, Italiano A.
\newblock Quality of reporting of phase II trials: A focus on highly ranked
  oncology journals.
\newblock \emph{Ann Oncol}  2014;25:536--41.

\bibitem{hanfelt1999}
Hanfelt J, Slack R, Gehan E.
\newblock A modification of Simon's optimal design for phase II trials when the
  criterion is median sample size.
\newblock \emph{Control Clin Trials}  1999;20:555--66.

\bibitem{jung2004a}
Jung S, Lee T, Kim K, George S.
\newblock Admissible two‐stage designs for phase II cancer clinical trials.
\newblock \emph{Stat Med}  2004;23:561--9.

\bibitem{mander2010}
Mander A, Thompson S.
\newblock Two-stage designs optimal under the alternative hypothesis for phase
  II cancer clinical trials.
\newblock \emph{Contemp Clin Trials}  2010;31:572--8.

\bibitem{clopper1934}
Clopper C, Pearson E.
\newblock The use of confidence or fiducial limits illustrated in the case of
  the binomial.
\newblock \emph{Biometrika}  1934;26:404--13.

\bibitem{porcher2012}
Porcher R, Desseaux K.
\newblock What inference for two-stage phase II trials?
\newblock \emph{BMC Med Res Methodol}  2012;12:117.

\bibitem{jung2006}
Jung S, Owzar K, George S, Lee T.
\newblock p-value calculation for multistage phase II cancer clinical trials.
\newblock \emph{J Biopharm Stat}  2006;16:765--75.

\bibitem{jennison1983}
Jennison C, Turnbull B.
\newblock Confidence intervals for a binomial parameter following a multistage
  test with application to MIL-STD 105D and medical trials.
\newblock \emph{Technometrics}  1983;25:49--58.

\bibitem{jung2004b}
Jung S, Kim K.
\newblock On the estimation of the binomial probability in multistage clinical
  trials.
\newblock \emph{Stat Med}  2004;23:881--96.

\bibitem{wu2008}
Wu Y, Shih W.
\newblock Approaches to handling data when a phase II trial deviates from the
  pre-specified Simon’s two-stage design.
\newblock \emph{Stat Med}  2008;27:6190--208.

\bibitem{khan2012}
Khan I, Sarker SJ, Hackshaw A.
\newblock Smaller sample sizes for phase II trials based on exact tests with
  actual error rates by trading-off their nominal levels of significance and
  power.
\newblock \emph{Br J Cancer}  2012;107:1801--9.

\bibitem{mander2012}
Mander A, Wason J, Sweeting M, Thompson S.
\newblock Admissible two-stage designs for phase II cancer clinical trials that
  incorporate the expected sample size under the alternative hypothesis.
\newblock \emph{Pharm Stat}  2012;11:91--6.

\bibitem{koyama2008}
Koyama T, Chen H.
\newblock Proper inference from Simon's two-stage designs.
\newblock \emph{Stat Med}  2008;27:3145--54.

\bibitem{turner2012}
Turner L, Shamseer L, Altman D, Schulz K, Moher D.
\newblock Does use of the CONSORT statement impact the completeness of
  reporting of randomised controlled trials published in medical journals? A
  Cochrane review.
\newblock \emph{Syst Rev}  2012;1:60.

\bibitem{guo2005}
Guo H, Liu A.
\newblock A simple and efficient bias-reduced estimator of response probability
  following a group sequential phase II trial.
\newblock \emph{J Biopharm Stat}  2005;15:773--81.

\bibitem{chang1989}
Chang M, Wieand H, Chang V.
\newblock The bias of the sample proportion following a group sequential phase
  II clinical trial.
\newblock \emph{Stat Med}  1989;8:563--70.

\bibitem{pepe2009}
Pepe M, Feng Z, Longton G, Koopmeiners J.
\newblock Conditional estimation of sensitivity and specificity from a phase 2
  biomarker study allowing early termination for futility.
\newblock \emph{Stat Med}  2009;28:762--79.

\bibitem{tsai2008}
Tsai W, Chi Y, Chen C.
\newblock Interval estimation of binomial proportion in clinical trials with a
  two-stage design.
\newblock \emph{Stat Med}  2008;27:15--35.

\bibitem{bryant1995}
Bryant J, Day R.
\newblock Incorporating toxicity considerations into the design of two-stage
  phase II clinical trials.
\newblock \emph{Biometrics}  1995;51:1372--83.

\bibitem{fleming1982}
Fleming T.
\newblock One-sample multiple testing procedure for phase II clinical trials.
\newblock \emph{Biometrics}  1982;38:143--51.

\end{thebibliography}

\newpage
\appendix

\section*{Supplementary materials to `Two-stage single-arm trials are rarely reported adequately'}

\beginsupplement
\renewcommand{\figurename}{Supplementary figure}
\renewcommand{\tablename}{Supplementary table}

\section*{Supplementary methods}

\subsection*{Two-stage group-sequential single-arm trial designs for a binary primary outcome}

In the main manuscript, we outlined the design of ``Simon two-stage" trials. Here, we elaborate on the statistical details of performing adjusted inference and handling design deviation in such studies.

First, let
\begin{align*}
    b(s | m, p) &= \binom{m}{s}p^s(1-p)^{m-s},\\
    B(s | m, p) &= \sum_{i=0}^sb(i | m,p),
\end{align*}
be the probability density and cumulative distribution function of a $Bin(m,p)$ random variable. Next, let $s_m=\sum_{i=1}^mx_i$ be the number of successes seen for the primary outcome in the first $m$ patients. Finally, set $\boldsymbol{a}=(a_1,a)$ and $\boldsymbol{n}=(n_1,n)$. Then, define $U_j(s | p,\boldsymbol{a},\boldsymbol{n})$ to be the probability that $s$ successes are observed by the end of stage $j$. We then have
\begin{align*}
    U_1(s | p,\boldsymbol{a},\boldsymbol{n}) &= b(s | n_1,p),\\
    U_2(s | p,\boldsymbol{a},\boldsymbol{n}) &= \sum_{i=\max\{a_1+1,s-(n-n_1)\}}^{\min(n_1,s)}b(i | n_1,p)b(s-i | n-n_1,p).
\end{align*}

Any point-estimation procedure, $\hat{P}$, must specify values for $\hat{p}$ for all possible numbers of successes that could be seen on trial termination at the end of each stage. We denote these values by $\hat{p}=\hat{p}(s,m)$, and then specifically we must have them for $(s,m)=(0,n_1),\dots,(a_1,n_1),(a_1+1,n),\dots,(n,n)$.

As discussed, a `naive' point estimation procedure, $\hat{P}_\text{naive}$, that does not account for the incorporated interim analysis, sets $\hat{p}_\text{naive}(s,m) = s/m$. In our work, we also consider six adjusted estimation procedures that have been proposed in the literature. For this, some final notation is useful: for any point estimation procedure $\hat{P}$, we can compute its expected value and bias through
\begin{align*}
    E(\hat{P} | p,\boldsymbol{a},\boldsymbol{n}) &= \sum_{s=0}^{a_1}\hat{p}(s,n_1)U_1\{s | p,\boldsymbol{a},\boldsymbol{n}\} + \sum_{s=a_1+1}^{n}\hat{p}(s,n)U_2\{s | p,\boldsymbol{a},\boldsymbol{n}\},\\
    Bias(\hat{P} | p,\boldsymbol{a},\boldsymbol{n}) &= E(\hat{P} | p,\boldsymbol{a},\boldsymbol{n}) - p.
\end{align*}

The considered adjusted point estimates are then
\begin{itemize}
    \item The bias-subtracted estimate\cite{guo2005}
    $$\hat{p}_\text{bias-sub}(s,m) = \hat{p}_\text{naive}(s,m) - Bias\{\hat{P}_\text{naive} | \hat{p}_\text{naive}(s,m),\boldsymbol{a},\boldsymbol{n}\}.$$
    \item The bias-adjusted estimate\cite{chang1989}, which is given by the numerical solution to
    $$\hat{p}_\text{bias-adj}(s,m) = \hat{p}_\text{naive}(s,m) - Bias\{\hat{P}_\text{naive} | \hat{p}_\text{bias-adj}(s,m),\boldsymbol{a},\boldsymbol{n}\}.$$
    \item The uniform minimum variance unbiased estimate\cite{jung2004b}
    $$\hat{p}_\text{umvue}(s,m) = \begin{cases} \frac{s}{n_1} &: m=n_1,\\ \frac{\sum_{i=\max\{a_1+1,s-(n-n_1)\}}^{\min(s,n_1)}\binom{n_1-1}{i-1}\binom{n-n_1}{s-i}}{\sum_{i=\max\{n_1+1,s-(n-n_1)\}}^{\min(s,n_1)}\binom{n_1}{i}\binom{n-n_1}{s-i}} &: m = n.\end{cases}$$
    \item The uniform minimum variance conditionally unbiased estimate\cite{pepe2009}
    $$\hat{p}_\text{umvcue}(s,m) = \begin{cases} \frac{s}{n_1} &: m=n_1,\\ \frac{\sum_{i=\max\{a_1+1,s-(n-n_1)\}}^{\min(s,n_1)}\binom{n_1}{i}\binom{n-n_1-1}{s-i-1}}{\sum_{i=\max\{n_1+1,s-(n-n_1)\}}^{\min(s,n_1)}\binom{n_1}{i}\binom{n-n_1}{s-i}} &: m = n.\end{cases}$$
    \item The conditional estimate\cite{tsai2008}, which sets $\hat{p}_\text{cond}(s,m)=s/m$ for $m=n_1$. When $m=n$, $\hat{p}_\text{cond}(s,n)$ is the numerical maximiser of
    $$-\log\left\{ \sum_{i=a_1 + 1}^{n_1}b(i | n_1,p) \right\} + s\log(p) + \{(n-n_1) - s\}\log(1-p).$$
    \item The median unbiased estimate\cite{koyama2008}, which sets $\hat{p}_\text{mue}(s,m)$ as the numerical solution to
    $$ q\{s,m | \hat{p}_\text{mue}(s,m)\}=0.5,$$
    where
    $$ q\{s,m | p \} = \begin{cases} 1 - B(s-1 | n_1,p) &: m=n_1,\\ \sum_{i=a_1+1}^{n_1} b(i | n_1,p)\{1 - B(s-i-1 | n-n_1,p)\} &: m=n,\end{cases}. $$
\end{itemize}

We also consider two adjusted confidence interval (CI) procedures and several unadjusted CI procedures. As for the point estimation procedures above, any CI procedure, $C$, must specify $c(s,m)=(p_\text{low}(s,m),p_\text{upp}(s,m))$ for all possible $(s,m)$ combinations on trial termination. The first considered adjusted CI procedure\cite{jennison1983} sets $p_\text{low}(s,m)$ and $p_\text{upp}(s,m)$ for a $100(1-\alpha)$\% CI as the numerical solutions to
\begin{align*}
    q\{s,m | p_\text{low}(s,m)\} &= 1-\alpha/2,\\
    q\{s,m | p_\text{upp}(s,m)\} &= \alpha/2.\\
\end{align*}

The second procedure\cite{porcher2012}, a mid-$p$ approach, sets
\begin{scriptsize}
\[ \sum_{\{(i,j) : \hat{p}_\text{umvue}(i,j) > \hat{p}_\text{umvue}(s,m)\}}U_j\{i|p_\text{low}(s,m),\boldsymbol{a},\boldsymbol{n}\} + 0.5\sum_{\{(i,j) : \hat{p}_\text{umvue}(i,j) = \hat{p}_\text{umvue}(s,m)\}}U_j\{i|p_\text{low}(s,m),\boldsymbol{a},\boldsymbol{n}\} = 1-\alpha/2, \]
\end{scriptsize}
and similarly for $p_\text{upp}(s,m)$.

We compare these to a number of standard unadjusted procedures, in particular Clopper-Pearson\cite{clopper1934}, which for $s\notin\{0,n\}$ sets
\begin{align*}
    p_\text{low}(s,m) &= \left[ 1 + \frac{m-s+1}{sF^{-1}\{\alpha/2,2s,2(m-s+1)\}} \right]^{-1},\\
    p_\text{upp}(s,m) &= \left[ 1 + \frac{m-s+1}{(s+1)F^{-1}\{1-\alpha/2,2(s+1),2(m-s)\}} \right]^{-1},\\
\end{align*}
where $F^{-1}(q,x,y)$ is the $100q$\% quantile of an $F$-distribution with $x$ and $y$ degrees of freedom. Supplementing this, $c(0,n_1) = (0,1-(\alpha/2)^{1/n_1})$ and $c(n,n) = ((\alpha/2)^{1/n},1)$.

For further information on the above, see Porcher and Desseaux\cite{porcher2012}.

Finally, we describe how a determination can be made on $H_0$ in the presence of design deviation. Englert and Kieser\cite{englert2015} provide comprehensive methodology for this by describing the decision rules for any design in terms of a discrete conditional error function. Whilst they allow for the interim analysis being conducted after any number of patients (i.e., a different value from that specified in the design), here we describe handling of deviation only in the timing of the stage two analysis, as just one paper noted that they conducted their interim analysis at an unplanned point. To determine whether to reject $H_0$ at the end of the second stage, if the analysis is conducted based on $n_\text{an}$ patients data ($n_\text{an}$ not necessarily equal to $n$), a $p$-value based on the second stage data only is first computed
$$p_2(s_{n_\text{an}}-s_{n_1},n_\text{an}-n_1) = 1 - B(s_{n_\text{an}}-s_{n_1}-1 | p_0,n_\text{an}-n_1).$$
Then, $H_0$ is rejected if $p_2(s_{n_\text{an}}-s_{n_1},n_\text{an}-n_1) \le D(s_{n_1}|\boldsymbol{a},\boldsymbol{n})$, for
$$ D(s|\boldsymbol{a},\boldsymbol{n}) = \begin{cases} 0 &: s \le a_1,\\ 1 - B(a-s | p,n-n_1) &: a_1 < s \le a,\\ 1 &: s > a.\end{cases} $$

\subsection*{Literature review}

\subsubsection*{Inclusion criteria}

As discussed in the main manuscript, PubMed was searched using the following term: (``2013/01/01"[Date - Publication] : ``2017/12/31"[Date - Publication]) AND Clinical Trial[Publication Type] AND (phase II[Title/Abstract] OR phase 2[Title/Abstract]) AND (cancer[All Fields] OR oncology[All Fields]). Thus, we reviewed clinical trial publications over a five year period, checking all such articles that included the phrase `phase II' or `phase 2' in the title/abstract and the word `cancer' or `oncology' in any field.

We wished to include as many treatment arms as possible, whilst not biasing our findings. Our initial inclusion criteria for arms reported in the returned records were therefore as follows:

\begin{itemize}
    \item Accessible full-length articles. No short communications were allowed as the quality of reporting would be expected to be lower in such articles. We required access to the complete text for inclusion.
    \item Primary publication on an arm's complete results. No secondary or preliminary analyses were allowed as the quality of reporting would be expected to be lower in such articles.
    \item Treatment arms designed and analysed using ``Simon's two-stage design". Thus
    \begin{itemize}
        \item Treatment arms that were designed using Simon's approach, but for which inference was performed as if the primary outcome was a time-to-event variable (e.g., a dichotomised version of PFS was used for design purposes, but inference was based on the exact PFS event-time data) were to be excluded. These were to be excluded as we planned to re-analyse included arms using appropriate adjusted methods for a binary primary outcome and compare our findings to the reported results; such comparisons would not be logical across analysis frameworks.
        \item Treatment arms using Bayesian methods were to be excluded, as analysis and reporting of such trials would be expected to differ (e.g., computation and reporting of adjusted point estimates would be expected to be less frequent).
        \item Treatment arms with $p_1<p_0$ (e.g., where the primary outcome is toxicity) were to be excluded. Though such trials could easily be analysed using adjusted procedures like those outlined above, we anticipated some authors may not be aware of this.
        \item Treatment arms for which the design was intentionally modified were to be excluded. Such a design would then correspond to an adaptive approach and not the considered group-sequential design.
    \end{itemize}
\end{itemize}

Five hundred and thirty four articles were randomly selected for evaluation for inclusion by MJG and APM based on the above criteria. The authors agreed on inclusion for 520 articles (97.3\%). The authors disagreed on the remaining 12 articles for six reasons, which led to the following additional clarifications on inclusion

\begin{itemize}
    \item Treatment arms stated to have used ``Simon's two-stage design" but for which no additional information was available to confirm this should be excluded (e.g., no reporting of stage-wise sample sizes or other discussion of interim analysis). We felt inclusion of such articles would bias downwards the identified quality of reporting and that presentation of what are likely conservative values in some instances would be more appropriate.
    \item Treatment arms for which the results of sequential trials are reported simultaneously should be excluded (e.g., simultaneous reporting of phase I and II results). We again felt inclusion of such articles would potentially bias our results, as substantial focus may not be given to the Simon-designed component. In some cases, it was also unclear as to whether data had been combined across phases and if re-analysis using the considered adjusted inference procedures would then be appropriate.
    \item Arms from articles reporting the results of multiple treatment arms should be included only if each arm was designed and analysed independently (e.g., the overarching design could not be a randomised selection approach, or similar). We anticipated these trials may focus more on arm differences and not on evaluating point estimates or confidence intervals for each arm separately.
    \item Treatment arms for which it was stated formal interim monitoring of toxicity was performed should be excluded. Such trials would be more similar to a Bryant and Day\cite{bryant1995} type approach than Simon's methodology.
    \item Treatment arms stated to have been designed using ``Fleming's" approach\cite{fleming1982} should be excluded, as we felt such trials may not report interim sample sizes or stopping rules as frequently given the way Fleming designs are identified.
    \item Treatment arms for which it was stated there was more than one primary outcome should be excluded. For one article it was not clear how a second primary outcome may have influenced design, analysis, and reporting. We therefore decided to exclude all subsequent articles with more than one primary outcome.
\end{itemize}

A decision was then taken that given there was by enlarge agreement on inclusion, the remaining articles were to be assessed for inclusion by MJG only, but that discussion with APM would be conducted if required for any individual article.

\subsubsection*{Data extraction}

For each of the 534 articles reviewed by both authors, a pilot duplicate data extraction was performed for the treatment arms they deemed to be eligible for inclusion. In total, data was extracted by both authors for 58 eligible treatment arms; in each case data was extracted for 28 questions. These were Questions 1, 4, 5, 7-13, 15, 18-21, and 23-34 from Table 1 in the main manuscript, along with ``What was the total number of efficacy evaluable patients?". Following the pilot evaluation, this last question was deemed to be an unsuitable method of extracting data that would facilitate trial re-analyses, as some studies did not utilise the number of evaluable patients in their analysis. This was therefore replaced by Question 22, ``If `Yes' to Q21, what was the sample size assumed in the analysis?". Following the pilot data extraction, the remaining questions listed in Table 1 (Questions 2, 3, 6, 14, 16, and 17) were also added to enhance our evaluation. Data for these additional questions was subsequently extracted by MJG for those arms included during the pilot.

We are thus able to evaluate the replicability of extraction across 58 treatment arms and 27 of the final questions used for our results. As discussed in the main manuscript, across 14 questions requiring non-binary value extraction (Questions 4, 5, 7-13, 20, 23, 26, 33, 34), the duplicate extractions agreed 96.2\% of the time. Across a wider set of 26 questions (all except Question 1, which we omit as equality for this is implied by performing a comparison), agreement occurred 94.3\% of the time. With replicability high, duplicate extraction was not performed for the remaining included arms.

Note that data was extracted from the included articles and their supplementary materials, but not from any linked protocols. This is because we believed all questions amounted to critical information that should be present in the primary publication reporting a trial's results.

\subsubsection*{Trial re-analyses}

For those trials that reported a point estimate or CI not stated to have been adjusted, the reported results were re-analysed were possible to evaluate consistency with unadjusted and adjusted procedures. We limited this re-analysis to those trials that terminated in stage two, for as can be seen above, point estimation and CI procedures typically only modify standard inferential methods if a trial proceeds to stage two. To this end, denote the number of realised patients assumed in the two analyses by $\boldsymbol{n}_\text{an}=(n_1,n_\text{an})$, where $n_\text{an}$ is that extracted in Question 22.

First, the reported point estimate was checked for consistency with $\hat{p}_\text{naive}$ to the reported number of decimal places. For example, suppose the data for Questions 22 and 23 indicate 21 patients were assumed in the analysis with 7 successes, and Question 25 indicates that the reported point estimate was 0.33. This would be consistent with $\hat{p}_\text{naive}$ to the reported number of decimal places, as $\hat{p}_\text{naive}=7/21=0.33$ to 2 decimal places. So long as $a_1$ and $n_1$ were also reported clearly (Questions 10 and 12), this point estimate was also evaluated for whether it was consistent with each of the six adjusted point estimates described earlier.

An equivalent approach was performed to evaluate whether any reported CI was consistent with one of four unadjusted (Clopper-Pearson, Wald, Wilson, and Blyth-Still-Casella; the four unadjusted CI methods stated to have been used by any included article across our review) and two adjusted CIs (those given earlier).

For those included treatment arms that reported a 95\% CI, we also compared the coverage provided by their CI procedure to the adjusted method of Jennison and Turnbull\cite{jennison1983} when the true success probability is $\hat{p}_\text{umvue}$. To this end, note that coverage can be computed for any $p$ and $C$ via
$$ Cover(C | p,\boldsymbol{a},\boldsymbol{n}_\text{an}) = \sum_{i=0}^{a_1}\mathbb{I}\{p\in c(s,n_1)\}U_1(s | p,\boldsymbol{a},\boldsymbol{n}_\text{an}) + \sum_{i=a_1}^{n}\mathbb{I}\{p\in c(s,n_\text{an})\}U_2(s | p,\boldsymbol{a},\boldsymbol{n}_\text{an}).$$

Finally, Figure 5 compares the unconditional operating characteristics for the sample sizes given in $\boldsymbol{n}_\text{an}$, when either $a$ is retained, or the method of Englert and Kieser\cite{englert2015} described earlier is implemented. For example, we compare the type-I error-rates via
\begin{align*}
  P(\text{Reject } H_0\text{ when }a\text{ is retained }|p_0,\boldsymbol{a},\boldsymbol{n}_\text{an}) &= \sum_{s=a+1}^{n_\text{an}}U_2(p_0,\boldsymbol{a},\boldsymbol{n}_\text{an}),\\
  P(\text{Reject } H_0\text{ when EK is used }|p_0,\boldsymbol{a},\boldsymbol{n}_\text{an}) &= \sum_{s=0}^{n_1}b(s|p_0,n_1)[\mathbb{P}\{P_2(p,n_\text{an}-n_1) \le D(s|\boldsymbol{a},\boldsymbol{n})\}],
\end{align*}
where $P_2(p_0,n_\text{an}-n_1)$ is the random variable denoting the distribution of the second stage $p$-value under the boundary of $H_0$ for a second stage sample size of $n_\text{an}-n_1$. Power is compared similarly.

\subsection*{Supplementary results}

\subsubsection*{Additional result interpretation approaches}

\begin{figure}
	\centering
	\includegraphics[width=0.75\textwidth]{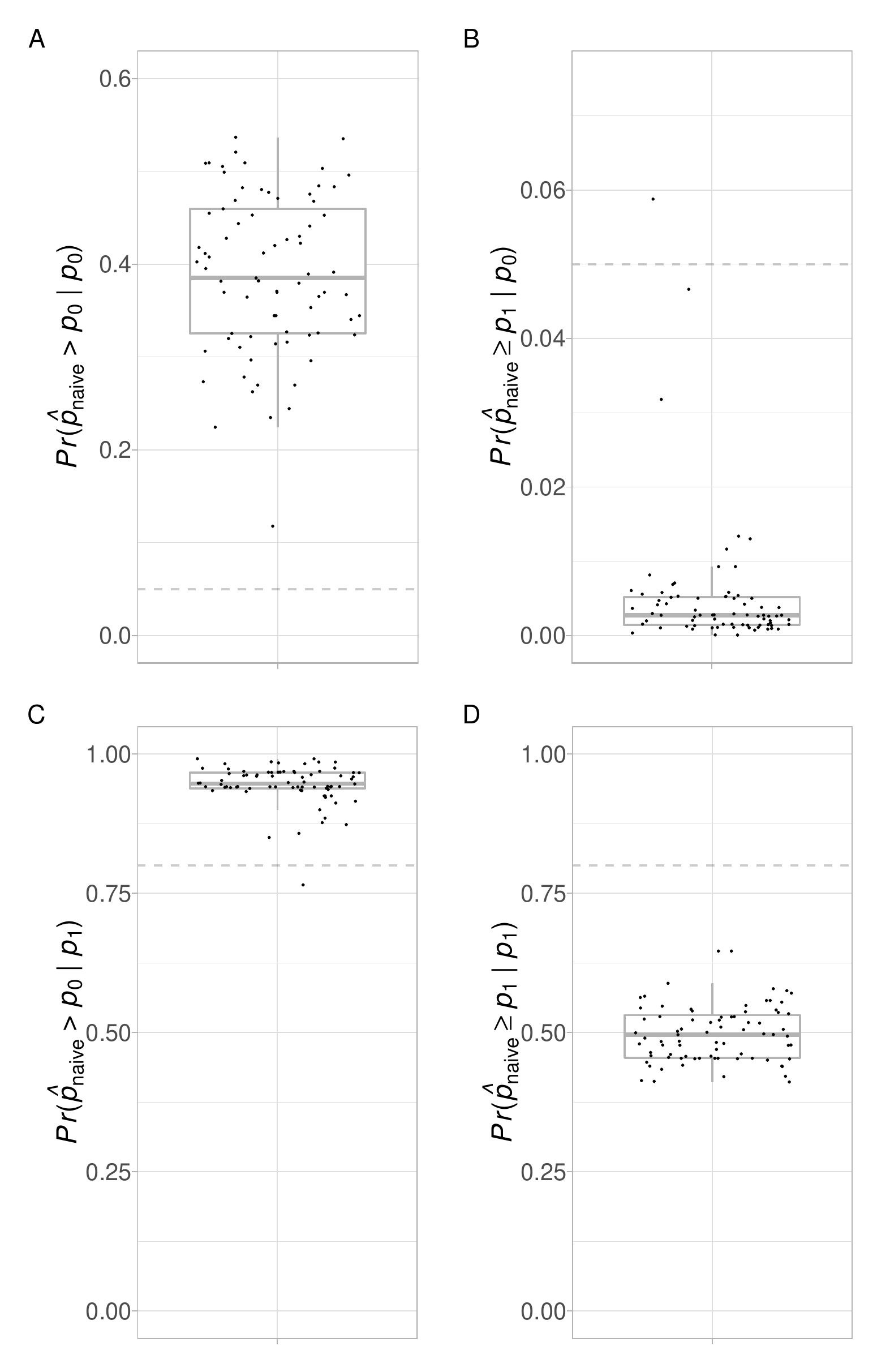}
	\caption{A comparison of the probabilities associated with several methods for interpreting the results of a two-stage single-arm trial. Values are given for 45 trials that terminated in stage 2 and specified $\alpha=0.05$ and $1-\beta=0.8$.\label{sfig1}}
\end{figure}

Extending Figure 5, Supplementary figure~\ref{sfig1} provides the probability of several quantities that trials were routinely observed to use in the interpretation of their results. These are: $\mathbb{P}(\hat{P}_\text{naive}>p_0 | p_0)$,  $\mathbb{P}(\hat{P}_\text{naive}>p_0 | p_1)$, $\mathbb{P}(\hat{P}_\text{naive}\ge p_1 | p_0)$, and $\mathbb{P}(\hat{P}_\text{naive}\ge p_1 | p_1)$. Note that, e.g., the first quantity is computed as
$$\mathbb{P}(\hat{P}_\text{naive}>p_0 | p_0) = \sum_{s=0}^{a_1}\mathbb{I}(\hat{p}_\text{naive}(s,n_1)>p_0)U_1(s|p_0,\boldsymbol{a},\boldsymbol{n}_\text{an}) + \sum_{s=a_1+1}^{n_\text{an}}\mathbb{I}(\hat{p}_\text{naive}(s,n_\text{an})>p_0)U_2(s|p_0,\boldsymbol{a},\boldsymbol{n}_\text{an}),$$
and similarly for the others.

Observe that the quantities conditioned on $p_0$ do not generally take values close to the target type-I error-rate 0.05. Similarly, the quantities conditioned on $p_1$ are not generally close to the target power 0.8.

\subsubsection*{Included articles that reported the results of more than one eligible treatment arm}

\begin{figure}
	\centering
	\includegraphics[width=0.75\textwidth]{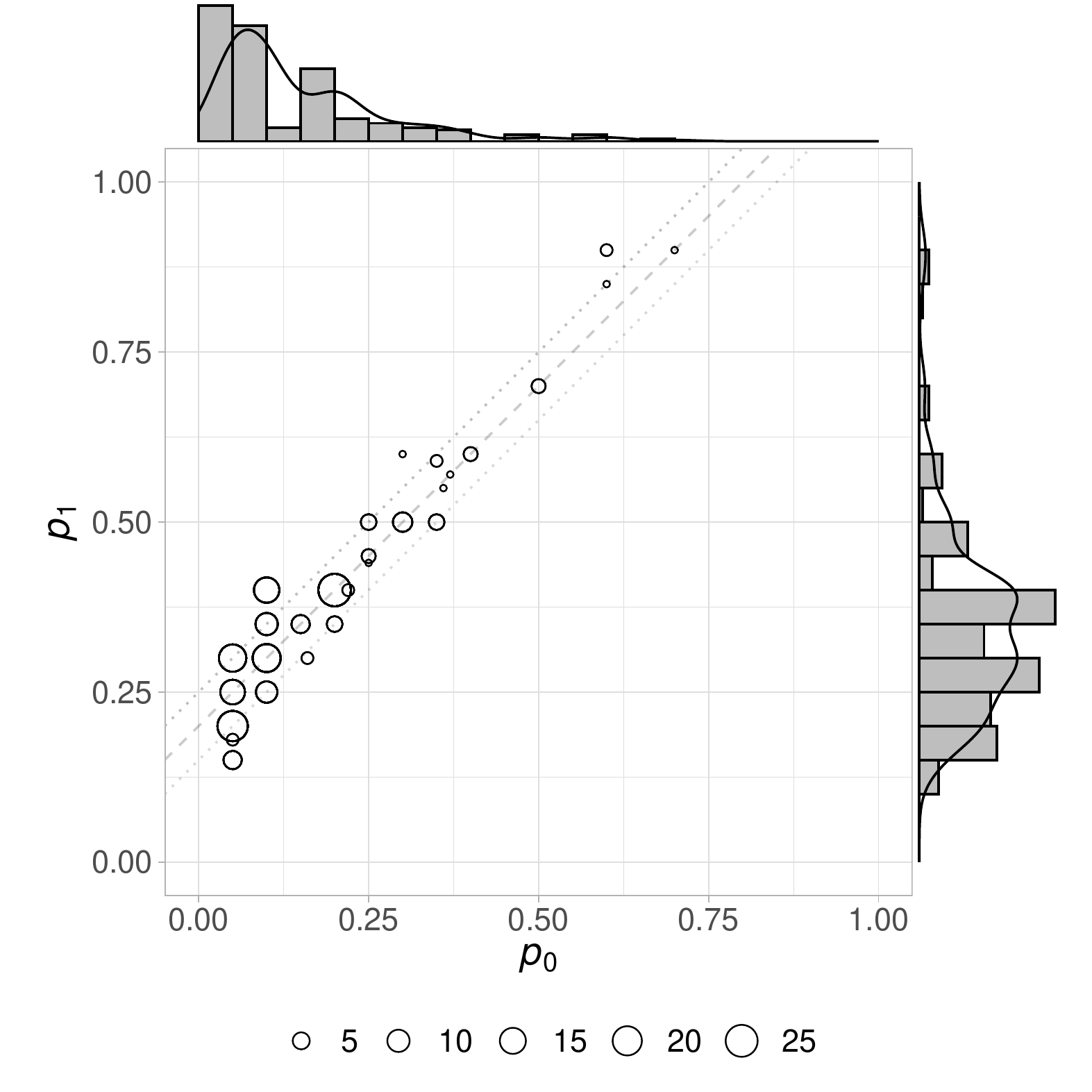}
	\caption{The combinations of $p_0$ and $p_1$ given for 185/204 arms (90.7\%) in the 75 included articles that reported the results of more than one eligible treatment arm. The size of the circles indicates the number of arms that had a particular combination. Densigrams are given on the axes to display the marginal distributions of $p_0$ and $p_1$. The dotted lines indicate the lower (0.15) and upper (0.25) quartiles of the distribution of $p_1-p_0$; the dashed line indicates the median (0.2) of this distribution.\label{sfig2}}
\end{figure}

\begin{figure}
	\centering
	\includegraphics[width=0.75\textwidth]{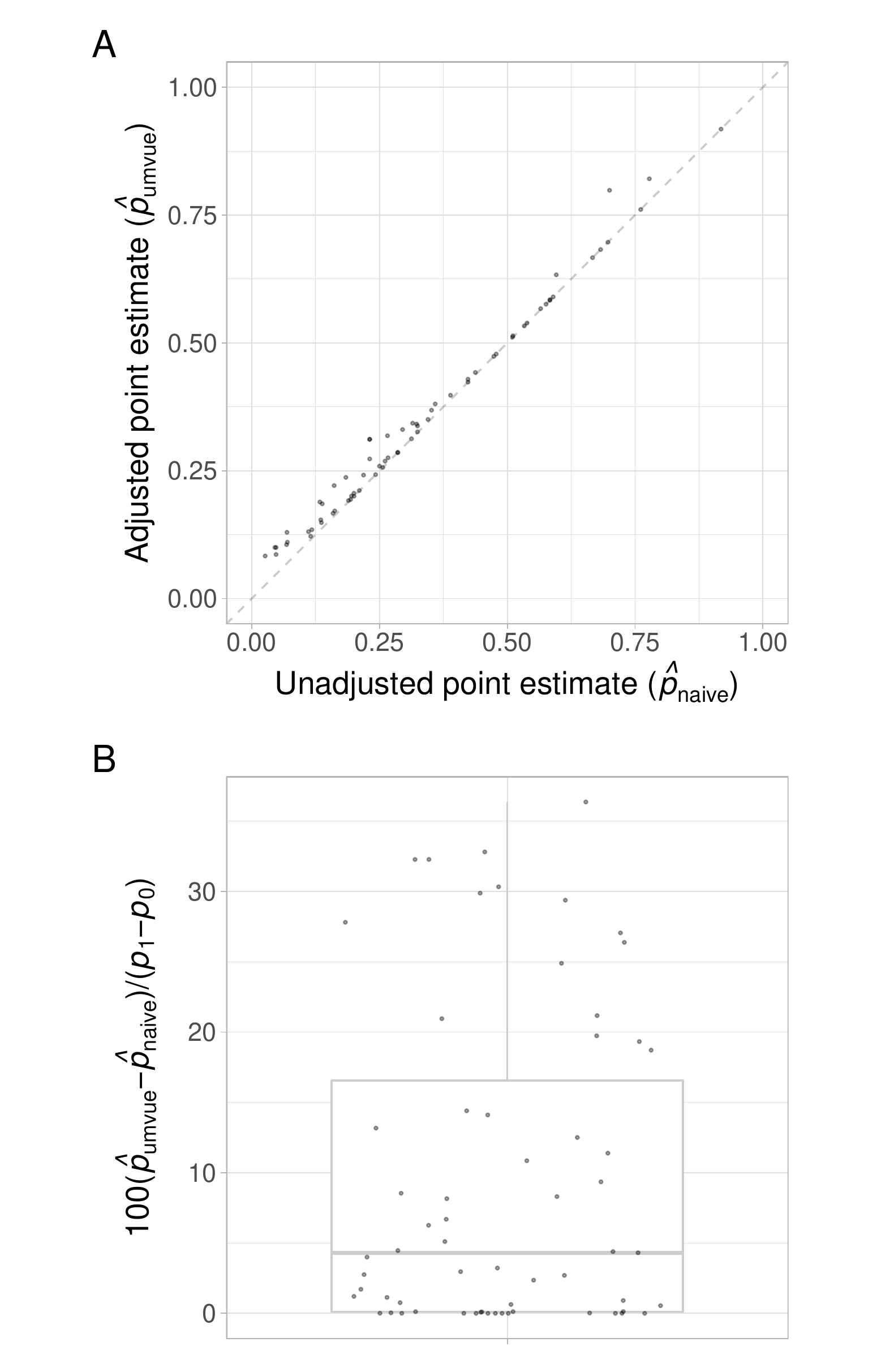}
	\caption{In A, a comparison of the naive unadjusted point estimate ($\hat{p}_\text{naive}$) and the uniform minimum variance unbiased estimate (UMVUE, $\hat{p}_\text{umvue}$) is given for 68 arms that terminated in stage 2 where the UMVUE could be computed. In B, the difference between $\hat{p}_\text{naive}$ and $\hat{p}_\text{umvue}$ is presented as a percentage of $p_1-p_0$, along with a boxplot to indicate the distribution of this data.\label{sfig3}}
\end{figure}

\begin{figure}
	\centering
	\includegraphics[width=0.75\textwidth]{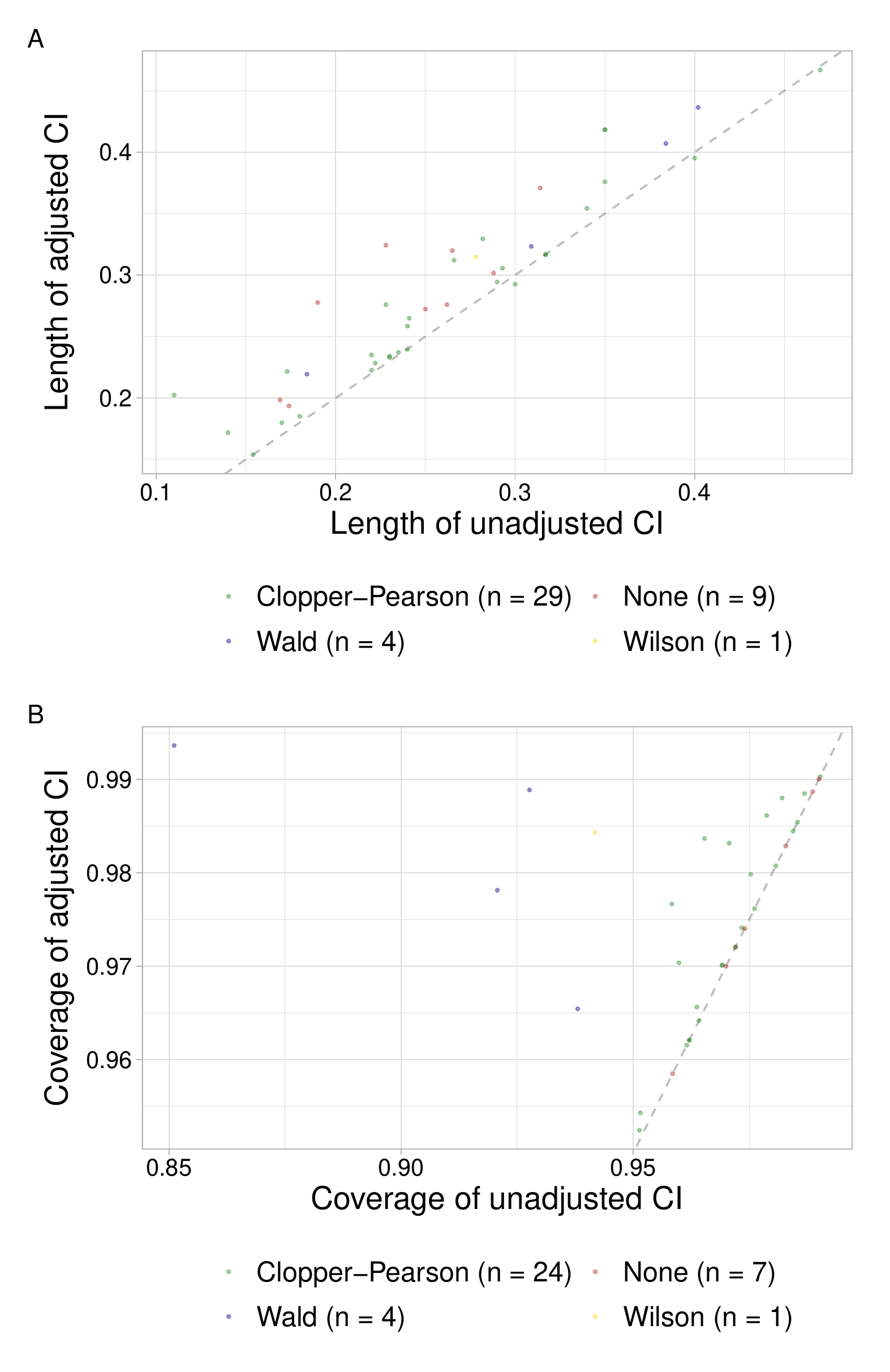}
	\caption{In A, the length of the reported unadjusted CI is compared to the length of the corresponding adjusted CI proposed by Jennison and Turnbull\cite{jennison1983} for the 43 arms for which this adjusted CI could be computed. In B, the respective coverage of these unadjusted and adjusted CIs when $p=\hat{p}_\text{umvue}$ is given for the 36 of these arms in which the target coverage was 0.95. In both cases, points are coloured by the unadjusted CI that the re-analysis indicated the reported CI matched with. For those CIs that matched none of the unadjusted CIs, Clopper-Pearson\cite{clopper1934} was used to compute the coverage.\label{sfig4}}
\end{figure}

\begin{figure}
	\centering
	\includegraphics[width=0.75\textwidth]{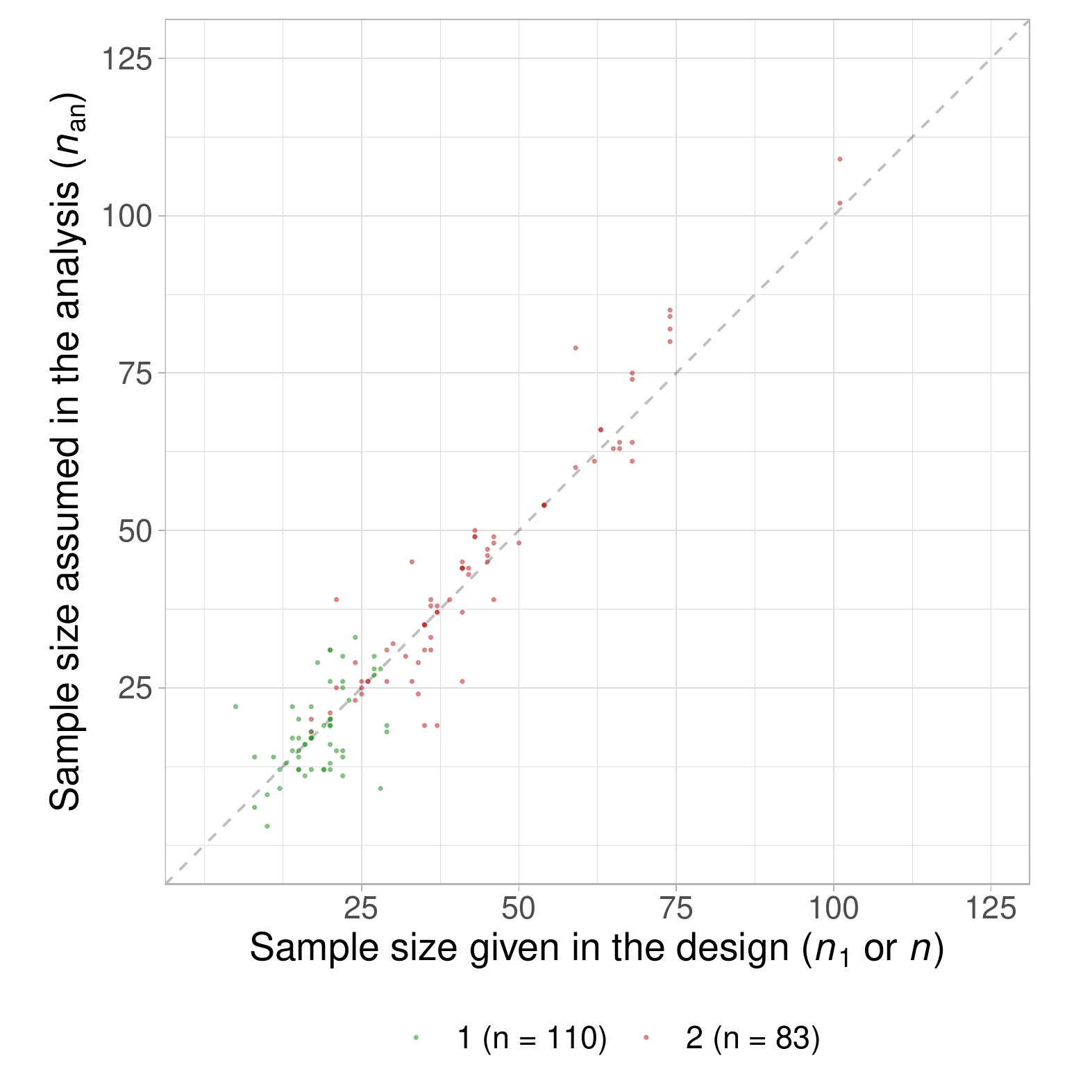}
	\caption{A comparison of the sample size required by the design and that assumed in the analysis for 193 articles. Colour indicates the judged stage of trial termination.\label{sfig5}}
\end{figure}

Supplementary 
Tables~\ref{stab:1}-\ref{stab:4} and Supplementary Figures~\ref{sfig2}-\ref{sfig5} display the corresponding results to those given in the main manuscript for the 204 treatment arms that were included across the 75 articles that reported the results of more than one eligible treatment arm.

The results are qualitatively similar to those presented in the main manuscript. In particular, for only 26.0\% of the 204 treatment arms were $p_0$, $p_1$, $\alpha$, $\beta$, the optimality criteria, $a_1$, $a$, $n_1$, and $n$ reported clearly. Adjusted point estimates (4 arms, 2.0\%) and adjusted confidence intervals (12 arms, 5.9\%) were again rarely reported, with little evidence to suggest that adjusted methods had been utilised with this unstated (Supplementary table~\ref{stab:4}).

Note that a larger proportion of the 204 treatment arms described here (53.9\%) were judged to have ended in stage one than for the 425 treatment arms described in the main manuscript (25.9\%). This is a consequence of many multi-arm trials struggling to recruit in treatment arms for rare conditions.

\begin{table}[htbp]
	\begin{center}
		\caption{Descriptors on the 204 treatment arms that were included across the 75 articles that reported the results of more than one eligible treatment arm. The denominators for computing percentages (given to 1 decimal place) are 75 and 204 respectively in all instances.\label{stab:1}}
		\begin{tabular}{llrr}
			\hline
			Descriptor & Value & By article & By arm\\
			 & & Number (\%) & Number (\%) \\
			\hline
			\rowcolor{Gray}
			Publication & 2013 & 21 (28.0) & 55 (27.0) \\
			\cellcolor{Gray}year & 2014 & 13 (17.3) & 36 (17.6) \\
			                 	\rowcolor{Gray}
			                 & 2015 & 23 (30.7) & 69 (33.8) \\
			\cellcolor{Gray}                 & 2016 & 12 (16.0) & 32 (15.7) \\
			                 	\rowcolor{Gray}
			                 & 2017 & 6 (8.0) & 12 (5.9)\\
			Journal & \textit{J Clin Onol} & 8 (10.7) & 29 (14.2) \\
			        & \cellcolor{Gray}\textit{Ann Oncol}                  & \cellcolor{Gray} 6 (8.0) & \cellcolor{Gray} 21 (10.3) \\
			        & \textit{Pediatr Blood Cancer}           & 3 (4.0) & 19 (9.3) \\
			        & \cellcolor{Gray}\textit{Eur J Cancer}                & \cellcolor{Gray} 5 (6.7) & \cellcolor{Gray} 13 (6.4) \\
			        & \textit{Lancet Oncol}                     & 6 (8.0) & 13 (6.4) \\
			        & \cellcolor{Gray}\textit{J Neurooncol}                 & \cellcolor{Gray} 4 (5.3) & \cellcolor{Gray} 11 (5.4) \\
			        & Other (27 journals)                 & 43 (57.3) & 98 (48.0) \\
		    \cellcolor{Gray}Cancer & \cellcolor{Gray}Lung       & \cellcolor{Gray}9 (12.0) & \cellcolor{Gray}31 (15.2) \\
			      \cellcolor{Gray} & Lymph      & 9 (12.0) & 21 (10.3) \\
			      \rowcolor{Gray} & Brain/CNS  & 7 (9.3) & 17 (8.3) \\
			      \cellcolor{Gray}  & Breast     & 7 (9.3) & 16 (7.8) \\
			      \rowcolor{Gray} & Blood      & 5 (6.7) & 14 (6.9) \\
			      \cellcolor{Gray} & Other      & 38 (50.7) & 105 (51.5) \\
			Stage of & \cellcolor{Gray}1 & \cellcolor{Gray} N/A & \cellcolor{Gray} 110 (53.9)\\
			termination & 2: Stated the criteria had been met for progression & N/A & 40 (19.6)\\
			& \cellcolor{Gray}2: Did not state the criteria had been met for progression & \cellcolor{Gray} N/A & \cellcolor{Gray} 43 (21.1)\\
			& Unclear & N/A & 11 (5.4)\\
			\hline
		\end{tabular}
	\end{center}
\end{table}

\begin{table}[htbp]
	\begin{center}
		\caption{Reporting of the design of the 204 treatment arms that were included across the 75 articles that reported the results of more than one eligible treatment arm. The denominator for computing percentages (given to 1 decimal place) is 204 in all instances.\label{stab:2}}
		\begin{tabular}{lr}
			\hline
			Criteria & Number (\%) \\
			\hline
			\rowcolor{Gray}
			Used the phrase ``Simon two-stage" (or similar) or cited Simon (1989)\cite{simon1989}               & 157 (77.0) \\
			Clearly stated $p_0$                                                                                & 192 (94.1) \\
			\rowcolor{Gray}
			Gave a justification for $p_0$                                                                      & 43 (21.1) \\
			\hspace{25pt}Citation given                                                                         & 32 (15.7) \\
			\rowcolor{Gray}
			\hspace{25pt}Justification given but no citation                                                    & 11 (5.4) \\
			Clearly stated $p_1$                                                                                & 191 (93.6) \\
			\rowcolor{Gray}
			Clearly stated $\alpha$                                                                             & 173 (84.8) \\
			\hspace{25pt}$\alpha=0.05$                                                                          & 57 (27.9) \\
			\rowcolor{Gray}
			\hspace{25pt}$\alpha=0.1$                                                                           & 76 (37.3) \\
			Clearly stated $\beta$                                                                              & 173 (84.8) \\
			\rowcolor{Gray}
			\hspace{25pt}$\beta=0.1$                                                                            & 95 (46.6) \\
			\hspace{25pt}$\beta=0.2$                                                                            & 40 (19.6) \\
			\rowcolor{Gray}
			Clearly stated the optimality criteria                                                              & 112 (54.9) \\
			\hspace{25pt}Null-optimal                                                                           & 69 (33.8) \\
			\rowcolor{Gray}
			\hspace{25pt}Minimax                                                                                & 41 (20.1) \\
			\hspace{25pt}Admissable                                                                             & 0 (0) \\
			\rowcolor{Gray}
			\hspace{25pt}Other                                                                                  & 2 (1.0) \\
			Clearly stated $a_1$                                                                                & 165 (80.9) \\
			\rowcolor{Gray}
			Clearly stated $a$                                                                                  & 130 (63.7) \\
			Clearly stated $n_1$                                                                                & 176 (86.3) \\
			\rowcolor{Gray}
			Clearly stated $n$                                                                                  & 163 (79.9) \\
			Indicated the recruitment target was greater than $n$                                               & 32 (15.7) \\
			\rowcolor{Gray}
			Clearly stated $p_0$ and $p_1$                                                                      & 185 (90.7) \\
			Clearly stated $p_0$, $p_1$, $\alpha$, and $\beta$                                                  & 164 (80.4) \\
			\rowcolor{Gray}
			Clearly stated $a_1$, $a$, $n_1$, and $n$                                                           & 130 (63.7) \\
			Clearly stated $p_0$, $p_1$, $\alpha$, $\beta$, and the optimality criteria                         & 86 (42.2) \\
			\rowcolor{Gray}
			Clearly stated $p_0$, $p_1$, $\alpha$, $\beta$, the optimality criteria, $a_1$, $a$, $n_1$, and $n$ & 53 (26.0) \\
			\hline
		\end{tabular}
	\end{center}
\end{table}

\begin{table}[htbp]
	\begin{center}
		\caption{Reporting of inferential procedures performed in the 204 treatment arms that were included across the 75 articles that reported the results of more than one eligible treatment arm, with additional stratification by stage of termination. The denominators for computing percentages (given to 1 decimal place) in the three columns are 110, 83, and 204 respectively unless stated otherwise.\label{stab:3}}
		\begin{tabular}{lrrr}
			\hline
			Criteria & Stage 1 & Stage 2 & All\\
			& Number (\%) & Number (\%) & Number (\%)\\
			\hline
			\rowcolor{Gray}
			Reported a point estimate, $p$-value, or confidence & 72 (65.5) & 81 (97.6) & 163 (79.9) \\
			\rowcolor{Gray}
			interval for the primary outcome & & &\\
			Reported a point estimate & 72 (65.5) & 81 (97.6) & 163 (79.9) \\
			\rowcolor{Gray}
			Stated the point estimate had been adjusted for & 0 (0) & 4 (4.8) & 4 (2.0) \\
			\rowcolor{Gray}
			the two-stage design & & &\\
			Reported a $p$-value                                                                      & 0 (0) & 0 (0) & 0 (0) \\
			\rowcolor{Gray}
			Stated the $p$-value had been adjusted for the & 0 (0) & 0 (0) & 0 (0) \\
			\rowcolor{Gray}
			two-stage design & & &\\
			Reported a confidence interval                                                          & 28 (25.5) & 57 (68.7) & 93 (45.6) \\
			\rowcolor{Gray}
			Stated the confidence interval had been & 1 (0.9) & 9 (10.8)  & 12 (5.9) \\
			\rowcolor{Gray}
			adjusted for the two-stage design & & &\\
			Analysis performed assuming a sample equal & 14/57 (24.6) & 13/73 (17.8) & 27/130 (20.8) \\
			to that given in the design & & &\\
			\hline
		\end{tabular}
	\end{center}
\end{table}

\begin{table}[htbp]
	\begin{center}
		\caption{Re-analysis of arms from articles that reported the results of more than one eligible treatment arm, which reported a point estimate or confidence interval not stated to have been adjusted. Consistency is measured in all cases against the reported number of decimal places. The denominators for computing percentages (given to 1 decimal place) are given in each row. Note that the re-analysis is limited to those articles that were judged to have terminated in stage 2.\label{stab:4}}
		\begin{tabular}{lrrr}
			\hline
			Criteria & Number (\%) \\
			\hline
			\rowcolor{Gray}
			Reported a point estimate not stated as adjusted and clearly reported the & 77/83 (92.8)\\
			\rowcolor{Gray}
			number of successes and sample size assumed in the analysis & \\
			Reported point estimate consistent with an unadjusted estimate & 75/77 (97.4) \\
			\rowcolor{Gray}
			Reported point estimate consistent with at least one adjusted estimate & 35/66 (53.0) \\
			Reported a confidence interval not stated as adjusted and clearly reported its &  \\
			level, the number of successes, and sample size assumed in the analysis & 48/84 (57.1) \\
			\rowcolor{Gray}
			Reported confidence interval consistent with at least one unadjusted interval & 37/48 (77.1) \\
			Reported confidence interval consistent with at least one adjusted interval & 2/44 (4.5) \\
			\hline
		\end{tabular}
	\end{center}
\end{table}

\end{document}